\documentclass[11pt, a4paper]{article}
\pdfoutput=1
\usepackage{jheparxiv}
\usepackage[latin1]{inputenc}
\usepackage{amsmath}
\usepackage{amsfonts}
\usepackage{amssymb}
\usepackage{latexsym}
\usepackage{mathrsfs}
\usepackage{graphicx}
\usepackage{color}
\usepackage{twistor2}
\usepackage[all]{xy}

\renewcommand{\d}{\mathrm{d}}

\subheader{\hfill \texttt{IHES/P/13/13}}

\title{Conformal and Einstein gravity from twistor actions}

\author{Tim Adamo}
\author{and Lionel Mason}

\affiliation{The Mathematical Institute\\
    University of Oxford \\
	24-29 St.~Giles', Oxford OX1 3LB \\
	United Kingdom}	
	
\emailAdd{adamo@maths.ox.ac.uk}
\emailAdd{lmason@maths.ox.ac.uk}

\date{\today}

\abstract{We use the embedding of Einstein gravity with cosmological constant into conformal gravity as a basis for using the twistor action for conformal gravity to obtain MHV scattering amplitudes not just for conformal gravity, but also for Einstein gravity on backgrounds with non-zero cosmological constant.  The new formulae for the gravitational MHV amplitude with cosmological constant arise by summing Feynman diagrams using the matrix-tree theorem.  We show that this formula is well-defined (i.e., is independent of certain gauge choices) and that it non-trivially reproduces Hodges' formula for the MHV amplitude in the flat-space limit.  We give a preliminary discussion of an MHV formalism for more general amplitudes obtained from the conformal gravity twistor action in an axial gauge.  We also see that the embedding of Einstein data into the conformal gravity action can be performed off-shell in twistor space to give a proposal for an Einstein twistor action that automatically gives the same MHV amplitude.  These ideas extend naturally to $\cN=4$ supersymmetry. }

\begin{document}

\maketitle


\section{Introduction}

Witten's twistor-string theory and related models \cite{Witten:2003nn, Berkovits:2004hg, Mason:2007zv} have inspired an extensive list of recent developments in our understanding of maximally supersymmetric ($\cN=4$) super-Yang-Mills (SYM) theory.  While these original twistor-string theories were limited in their applicability to perturbative gauge theory due to unwanted contributions from conformal gravity \cite{Berkovits:2004jj}, twistor actions for Yang-Mills theory were discovered which isolated the gauge theoretic degrees of freedom \cite{Mason:2005zm, Boels:2006ir}.  In one gauge they reduce directly to the space-time action.  However, in an axial gauge on twistor space,  twistor actions lead to particularly efficient Feynman diagrams, the MHV formalism, that had originally been suggested from twistor-string considerations \cite{Cachazo:2004kj}. In this formalism, the Maximal-Helicity-Violating (MHV) amplitudes are extended off-shell to provide vertices of a Feynman diagram-like formalism that are tied together with massless scalar propagators.  The MHV formalism successfully computes tree and  loop amplitudes at least in supersymmetric gauge theories \cite{Brandhuber:2004yw}.  Twistor actions have now been applied to study a wide variety of physical observables in $\cN=4$ SYM, including scattering amplitudes, null polygonal Wilson loops, and correlation functions (c.f., \cite{Adamo:2011pv} for a review).

There is also a twistor action for \emph{conformal gravity} \cite{Mason:2005zm}, the conformally invariant theory of gravity whose Lagrangian is the square of the Weyl tensor.  It has fourth-order equations of motion so its quantum theory is
non-unitary and is widely believed to be un-suitable for a physical theory.  Nevertheless, conformal gravity has
many interesting mathematical properties: for instance, it can be extended to supersymmetric theories for $\cN\leq 4$, and the maximally
supersymmetric theory ($\cN=4$) comes in several variants, some of which are
finite and power-counting renormalizable (c.f., \cite{Fradkin:1985am}
for a review).  More importantly for this paper, not only do solutions to Einstein gravity form a subsector of solutions to the field equations of conformal gravity,  Maldacena has shown that evaluated on a de Sitter background, the tree-level S-matrix for conformal gravity reduces to that for Einstein gravity when Einstein states are inserted \cite{Maldacena:2011mk}.  

Here we study the Einstein tree-level S-matrix by reducing the
conformal gravity twistor action to Einstein scattering data.
Most of our analysis concerns the reduction on-shell as in the
Maldacena argument, but in fact it can be done off-shell also.  This
off-shell reduction leads to a proposal for the twistor action for
general relativity itself; up to now there has only been a twistor
action for the self-dual sector of Einstein (super-) gravity
\cite{Mason:2007ct}.   In any case, the Maldacena argument shows that
the on-shell reduction of the full conformal gravity twistor action
will give the correct tree-level Einstein S-matrix and that is our
main focus here as, at this stage we do not have a direct classical
off-shell equivalence between the Einstein twistor action and the
standard Einstein action.  

Our main focus in this paper is on the MHV amplitude and both twistor actions by construction give rise to the same formulae for this.  In an axial gauge the twistor actions simplify considerably for the MHV amplitude, although they are by no means as simple as they are in the Yang-Mills case.  
We show that the Feynman diagrams for the MHV amplitudes can
nevertheless be summed using the matrix-tree theorem to give new forms
of the gravitational MHV amplitudes valid on backgrounds with
cosmological constant.  However, the transform of these into momentum
space is no longer straightforward because the de Sitter group does
not lead to ordinary momentum eigenstates.  Nevertheless, these ideas
provide an origin for the tree formula for the MHV amplitude of
\cite{Nguyen:2009jk} in terms of conventional Feynman diagrams and
their summation using the matrix-tree theorem as in
\cite{Feng:2012sy,Adamo:2012xe}. 

Progress on understanding the  amplitudes of (super-)gravity in twistor space has been much slower than for Yang-Mills--even at tree-level and at MHV.  In \cite{Mason:2008jy}  a form of the momentum space BGK formula for the MHV amplitude was proved by reformulating a background coupled scattering problem for a negative helicity particle on a self-dual background into twistor space.  
In \cite{Mason:2009sa} BCFW recursion was reformulated in twistor space to give formulae for gravity amplitudes that have similar support to their Yang-Mills counterparts. The MHV amplitude was shown to be obtainable as a sum of tree diagrams arising from BCFW recursion in \cite{Nguyen:2009jk}.  Progress was much more rapid after Hodges' discovery of a manifestly permutation invariant and compact formula for the MHV tree amplitude of Einstein gravity \cite{Hodges:2012ym}.  This was related to the tree-diagrams of \cite{Nguyen:2009jk} via the matrix-tree theorem in \cite{Feng:2012sy}.  It was generalized to the Cachazo-Skinner expression for the entire tree-level S-matrix of $\cN=8$ supergravity in terms of an integral over holomorphic maps from the Riemann sphere into twistor space \cite{Cachazo:2012kg, Cachazo:2012pz}.  Perhaps most striking is Skinner's development of a new twistor-string theory for $\cN=8$ Einstein supergravity \cite{Skinner:2013xp} that produces this formula. 

Parallel work sought to derive Einstein amplitudes from Witten and Berkovits' original twistor-string formula for conformal gravity amplitudes by restricting to Einstein states and appealing to Maldacena's argument to obtain the amplitudes \cite{Adamo:2012nn, Adamo:2012xe}.  Although the correct amplitudes are obtained at three points, the relationship between Einstein and conformal gravity amplitudes requires \emph{minimal} conformal supergravity rather than the non-minimal version arising from the Berkovits-Witten twistor-string (see \S\ref{minimal} for a discussion of this distinction).  Nevertheless, in \cite{Adamo:2012xe} it was shown that for MHV amplitudes, the correct Hodges formula is obtained at $n$-points when a tree ansatz is imposed on the worldsheet correlation function of the Berkovits-Witten twistor-string.  
Although there is no clear motivation for the tree ansatz within Witten's and Berkovits' twistor-string theory,\footnote{In Skinner's $\cN=8$ twistor-string the tree ansatz can be understood as arising from cancellation of the loops due to worldsheet supersymmetry.} it \emph{is} natural in the context of the Maldacena argument applied to the twistor action for conformal gravity \cite{Mason:2005zm}, which does give the minimal theory.  One aim of this paper is to give a complete presentation of that argument.  It also allows us to provide a generalization of the Hodges formula for the gravitational MHV amplitude to the case of non-vanishing cosmological constant, which is the regime where the Maldacena argument is most straightforwardly applicable.  We will see that this also arises from a twistor action for Einstein gravity, obtained by reducing the conformal gravity action to Einstein degrees of freedom. 

\emph{A priori} one could hope to derive a formula for the gravitational MHV amplitude with cosmological constant from Skinner's $\cN=8$ twistor-string \cite{Skinner:2013xp} as it is formulated for all values of $\Lambda$.  Unfortunately, although this twistor-string theory has been shown to give the correct tree-level amplitudes at $\Lambda=0$, it has so far not been possible to make sense of the worldsheet correlations functions for the $\Lambda\neq 0$ regime.  It is to be hoped that knowing the answer \eqref{NewAmpForm} will also allow us to understand how to properly understand this $\cN=8$  twistor-string for $\Lambda\neq 0$.

For a cosmological constant $\Lambda\neq 0$, the traditional definition of a scattering amplitude for asymptotically flat space-times no longer applies.  When $\Lambda >0$, one can still define mathematical quantities corresponding to scattering from past infinity to future infinity, but these are not \emph{physical} observables because no observer has access to the whole space-time.  These mathematical analogues of scattering amplitudes have become known as meta-observables \cite{Witten:2001kn}: the theory allows them to be computed, even if no single physical observer can ever measure them.  Actual physical observables can still be given in terms of the in-in formalism, where the observer only integrates over the portion of space-time containing his or her history.  When $\Lambda<0$, this situation is improved and the natural objects to compute are correlation functions in the conformal field theory on the boundary via the AdS/CFT correspondence (although mathematically the integration regions are not so dissimilar and indeed the formulae will be polynomial in $\Lambda$ so that the analytic continuation from positive to negative $\Lambda$ is trivial).  Our formulae for the MHV amplitude  (see \eqref{NewAmpForm} below) will essentially be in the form of an integrand that can be integrated either over all of space-time, as appropriate for the meta-observable for scattering from past to future infinity of de Sitter space, or over a different contour as required for the in-in formalism.
For the remainder of this paper, we will refer only to `scattering amplitudes' in de Sitter space, trusting the reader to keep the implicit subtleties in mind.  Furthermore, although we will focus on the case of $\Lambda >0$ de Sitter space in this paper, most of our arguments (and certainly the final formula) apply to anti-de Sitter space with trivial changes of sign and can be applied to the AdS/CFT correspondence.  

\smallskip 

We begin in Section \ref{CEG} with a discussion of the reduction of conformal gravity to Einstein gravity.  This includes a brief overview of different space-time action principles for conformal gravity and the relationship with general relativity on an asymptotically de Sitter background.  Maldacena's argument then indicates that the tree-level S-matrix of Einstein gravity can be computed from that of conformal gravity by restricting to Einstein degrees of freedom.  From this, we derive a precise version of the correspondence for space-time generating functionals of MHV amplitudes in Appendix \ref{GenFuncs}.  Since we will be interested in scattering amplitudes, we also discuss the relationship between polarization states for conformal and Einstein gravity.  These ideas also have a natural extension to maximally supersymmetric $\cN=4$ conformal supergravity, albeit only in the \emph{minimal} form as discussed in detail in \S\ref{minimal}.  In \S\ref{GenFuncs} we explain how the MHV amplitude can be constructed by scattering a negative helicity particle on a self-dual background, and how the embedding of Einstein gravity within conformal gravity works in this context.

In Section \ref{TAM}, we study the twistor action for minimal $\cN=4$ conformal supergravity.  After a brief review of some relevant aspects of twistor theory, we recall the definition of the twistor action for $\cN=0$ conformal gravity, and argue that its straightforward generalization to $\cN=4$ produces the minimal supersymmetric theory.  In this $\cN=4$ context, the data of conformal gravity is encoded by  a vector-valued $(0,1)$-form $f=f^{I}\partial_{I}$ on twistor space ($\CPT$) determining the deformation of the $\dbar$-operator away from some choice of background, and a $(1,1)$-form $g=g_{I}\d Z^{I}$, where $I$ is a super-twistor index.  The twistor action takes the form
\be{ICGTA}
S^{\mathrm{CG}}[g,f]=\int_{\CPT}\D^{3|4}Z\wedge g_{I}\wedge\left(\dbar f^{I}+[f,f]^{I}\right) - \varepsilon^{2}\int_{M}\d^{4|8}x\;\left[ \int_XY\cdot (\dbar_\sigma Z- f)+\left(\int_{X}g\right)^{2}\right],
\ee  
where $\D^{3|4}Z$ is a top degree form that is holomorphic with respect to the background complex structure on twistor space, $\varepsilon$ is the dimensionless coupling constant of conformal gravity, and $\d^{4|8}x$ is a canonically defined measure on the space of rational curves $X\subset\CPT$ corresponding to real points in space-time.   The first part is a local action on twistor space for the self-dual sector, and the third provides the required interactions to extend this to the full theory.  It includes an integral over the $4|8$ parameter family of rational curves $X$, which are holomorphic with respect to the almost complex structure and real with respect to a choice of reality structure. These are determined by the the second part, a Lagrange multiplier action $\int \rd^{4|8}x \, \int_X Y\cdot(\dbar_\sigma Z(x,\sigma) + f(Z) )$ for the almost complex curves $X$.

This conformal gravity twistor action is then restricted to Einstein degrees of freedom in Section \ref{TAM2}.  This leads to a twistorial expression for the generating functional of MHV amplitudes in Einstein (super-)gravity, and also results in a proposal for a twistor action describing Einstein gravity itself.  Under this reduction, the data for conformal gravity is replaced by a pair of $(0,1)$-forms $\tilde{h}$, $h$ of weights $-2$, $+2$ respectively (corresponding to the $\pm 2$ graviton multiplets for $\cN=4$), and twistor space is equipped with complex  weighted Poisson and contact structures, $\{\cdot, \cdot\}$ (of weight $-2$) and $\tau$ (a $(1,0)$-form of weight $+2$) respectively.  The reduction to a proposed twistor action for Einstein gravity is then:
\begin{multline}\label{IGRTA}
S^{\mathrm{Ein}}[\tilde{h},h]=\int_{\CPT}\D^{3|4}Z\wedge\tilde{h}\wedge\left(\dbar h+\frac{1}{2}\{h,h\}\right)-\\ \frac{\varepsilon^{2}}{\Lambda \kappa^{2}}\int_{M}\d^{4|8}x\;\left[ \int_XY\cdot (\dbar_\sigma Z+ \{h,Z\})+\left(\int_{X}\tilde{h}\wedge\tau\right)^{2}\right],
\end{multline}
where $\kappa^{2}=16\pi G_{N}$ and $\Lambda$ is the cosmological constant.  

We consider the Feynman diagrams for these actions in section \ref{MACC}.  In an axial gauge there are propagators on twistor space coming from the first part of the action and one on the Riemann sphere from the second.  For the MHV amplitude, only the Riemann sphere propagator enters and the calculation is the same for both actions.  These $\CP^1$ Feynman diagrams leads to a sum of tree diagrams for computing the Einstein amplitude within minimal $\cN=4$ conformal supergravity, or equivalently via the candidate Einstein action \eqref{IGRTA}.  This turns out to be essentially equivalent to the approach based on the Berkovits-Witten twistor-string in \cite{Adamo:2012xe} to obtain Hodges' formula from the \emph{non-minimal} conformal supergravity subject to a tree ansatze on the correlator; hence, we see that the tree ansatz  successfully isolates the minimal sector in the twistor-string (at MHV). 

By use of the matrix-tree theorem (which has already been utilized in the study of the flat-space MHV amplitude \cite{Nguyen:2009jk, Feng:2012sy}), we sum the Feynman diagrams to derive an expression for the MHV amplitude in the presence of a cosmological constant.  We find
\be{NewAmpForm}
\cM_{n,0}=\frac{1}{\Lambda}\int\frac{\d^{8|8}X}{\mathrm{vol}\;\GL(2,\C)}\left[ (X^2)^2\left|\HH^{12}_{12}\right| +\sum_{i,j,k,l}\omega^{1}_{ij}\omega^{2}_{kl}\left|\HH^{12ijkl}_{12ijkl}\right|\right]\prod_{m=1}^{n}h(Z(\sigma_m))\;\D\sigma_{m}\: ,
\ee
where $X^{I}_{A}$ modulo GL$(2,\C)$ are coordinates on the moduli space of degree one holomorphic maps $Z(\sigma)$ from $\CP^{1}$ (with homogeneous coordinates $\sigma^{A}$) to twistor space; $h(Z(\sigma_{i}))$ are twistor representatives for the external states; $\HH$ is the Hodges matrix\footnote{Here, and throughout the paper, we denote the $\SL(2,\C)$-invariant inner product on $\CP^{1}$ coordinates by $(ij)=\epsilon_{AB}\sigma^{A}_{i}\sigma^{B}_{j}$.  The notation $[\; , ]$ stands for a contraction with a skew bi-twistor $I^{IJ}$ called the \emph{infinity twistor} which is introduced in Section \ref{TAM2}.  Similarly, $\la\; , \ra$ denotes a contraction with the inverse infinity twistor $I_{IJ}$.}
\begin{equation*}
\HH_{ij}=\left\{
\begin{array}{c}
\frac{1}{(ij)}\left[\frac{\partial}{\partial Z(\sigma_{i})}, \frac{\partial}{\partial Z(\sigma_{i})}\right] \:\mbox{if}\: i\neq j \\
-\sum_{k\neq i}\HH_{ik}\frac{(\xi k)^{2}}{(\xi i)^2} \:\mbox{if}\: i=j
\end{array}\right. ,
\end{equation*}
expressed on twistor space extended to include a non-zero cosmological constant $\Lambda$, and the quantities $\omega^{1}_{ij}$ are given by
\begin{equation*}
\omega^{1}_{ij}=-\Lambda\frac{(1\xi)^{4}(ij)}{(1i)^{2}(1j)^{2}(\xi i)^{2}(\xi j)^{2}}\left[\frac{\partial}{\partial Z(\sigma_{i})},\frac{\partial}{\partial Z(\sigma_{j})}\right].
\end{equation*}
The notation $| \HH^{12}_{12}|$ indicates the determinant of $\HH$ with the row and columns corresponding to $h(Z(\sigma_1))$ and $h(Z(\sigma_2))$ removed, and $\xi\in\CP^{1}$ is an arbitrary reference spinor.  We prove that \eqref{NewAmpForm} is independent of the choice of $\xi$, and limits smoothly onto Hodges' formula when $\Lambda\rightarrow 0$.  This is non-trivial as the latter formula is the generalized determinant of a rank $(n-3)$ matrix whereas our original formula is a rank $n-2$ determinant; we show how our formula can be manipulated into such a form, providing the natural generalization of Hodges' formula to $\Lambda\neq 0$.

Section \ref{DC} concludes with a discussion of future directions following on from this work.  Most enticing is the possibility that the twistor actions studied here could be used to define a MHV formalism \cite{Cachazo:2004kj} for conformal gravity, and in turn Einstein gravity.  Indeed, the twistor action approach for $\cN=4$ SYM is one way of deriving this formalism in the gauge theory setting \cite{Boels:2007qn, Adamo:2011cb} where other techniques such as Risager recursion fail in the gravitational context \cite{BjerrumBohr:2005jr, Bianchi:2008pu}.  We also discuss how the twistor formula \eqref{NewAmpForm} could be converted into a meaningful physical observable in de Sitter space, as well as its potential relationship with twistor-string theory and background-coupled calculations.  The Appendices provide technical details for our arguments, as well as related background material that is not essential to the main results.

\medskip

\subsubsection*{\textit{Notation}}

Throughout this paper, we use the following index conventions: space-time tensor indices are Greek letters from the middle of the alphabet ($\mu,\nu=0,\cdots,3$); positive and negative chirality Weyl spinor indices are primed and un-primed capital Roman letters respectively ($A',B'=0',1'$ or $A,B=0,1$); $R$-symmetry indices are lower-case Roman indices from the beginning of the alphabet ($a,b=1,\ldots,\cN$).  We will also use bosonic twistor indices, denoted by Greek letters from the beginning of the alphabet ($\alpha,\beta$), as well as supersymmetric twistor indices, denoted by capital Roman letters from the middle of the alphabet ($I,J$).  

We denote the space of smooth $n$-forms on a manifold $M$ by $\Omega^{n}_{M}$; in the presence of a complex structure we denote the space of smooth $(p,q)$ forms by $\Omega^{p,q}_{M}$.  If we want to consider these spaces twisted by some sheaf $V$, then we write $\Omega^{n}_{M}(V)$ for `the space of smooth $n$-forms on $M$ with values in $V$,' and so forth.  Dolbeault cohomology groups on $M$ with values in $V$ are denoted by $H^{p,q}(M,V)$.  The complex line bundles $\cO(k)$ denote the bundles of functions homogeneous of weight $k$ on a (projective) manifold, and we make use of the abbreviation $\Omega^{n}_{M}(\cO(k))\equiv\Omega^{n}_{M}(k)$, and so on.  


\section{Einstein and Conformal Gravity on Asymptotically de Sitter Spaces}
\label{CEG}

This section is concerned with the space-time formulation of the various equations and structures that we will be dealing with.  The next section will carry on with the discussion in twistor space.

In the first instance we will work on 4-dimensional space-times $M$ with Lorentzian signature $(1,3)$ metrics $g$ that are asymptotically de Sitter.  By this, we shall mean that $g$ is complete and $M\simeq S^3\times (0,1)$ with smooth conformal compactification $(\bar M,\bar g)$ for $\bar M\simeq S^3\times [0,1]$ and $\bar g=\Omega^2 g$.   Here $\Omega \in C^\infty (\bar M)$ is the conformal factor, with $\Omega\neq 0$ on $M$, but $\Omega =0$, $\bar g(\rd \Omega,\rd \Omega) > 0$ at $\p \bar M$.  The geometry of the conformally flat case is described in detail in Appendix \ref{desitter-geom}.  In this context, the conformal gravity equations can be applied equally  to $g$ or $\bar g$.  We will see in this section that the Einstein equations imply those of conformal gravity and that the tree-level S-matrix for Einstein gravity in this context can be computed by use of that for conformal gravity.  This embedding extends to the $\cN=4$ minimal supersymmetric  extensions of these ideas and this is discussed on space-time in Appendix \ref{minimal} and in twistor space in the next section, where the extension is easiest to present.  Finally, as far as space-time considerations are concerned, in Appendix \ref{GenFuncs} we show how the MHV amplitude can be obtained by considering the scattering of an anti-self dual linear field on a nonlinear self-dual background and how the embedding of Einstein gravity into conformal gravity relates the generating functionals in this context.

\subsection{Conformal gravity}
 Conformal gravity is the theory obtained from the action
\be{CGA1}
S^{\mathrm{CG}}[g]=\frac{1}{\varepsilon^{2}}\int_{M}\d\mu\;C^{\mu\nu\rho\sigma}C_{\mu\nu\rho\sigma}=\frac{1}{\varepsilon^{2}}\int_{M}\d\mu\left(\Psi^{ABCD}\Psi_{ABCD}+\widetilde{\Psi}^{A'B'C'D'}\widetilde{\Psi}_{A'B'C'D'}\right),
\ee
where $\varepsilon^{2}$ is a dimensionless coupling constant, $\d\mu=\d^{4}x\sqrt{|g|}$ is the volume element, $C_{\mu\nu\rho\sigma}$ is the Weyl curvature tensor of $g$, and $\Psi_{ABCD}$, $\widetilde{\Psi}_{A'B'C'D'}$ are the anti-self-dual (ASD) and self-dual (SD) Weyl spinors respectively \cite{Penrose:1984}. This theory is conformally invariant and hence only depends upon (and constrains) the conformal structure $[g]$ underlying $g$.  The field equations are the vanishing of the Bach tensor, $B_{\mu\nu}$, which can be written in a variety of different forms thanks to the Bianchi identities:
\begin{eqnarray}\label{Bach}
B_{\mu\nu}&=&2\nabla^{\rho}\nabla^{\sigma}C_{\rho\mu\nu\sigma}+C_{\rho\mu\nu\sigma}R^{\rho\sigma} \nonumber \\
&=&\left(2\nabla_{\rho}\nabla_{(\mu}R^{\rho}_{\nu)}-\Box R_{\mu\nu}-\frac{2}{3}\nabla_{\mu}\nabla_{\nu}R
-2R_{\rho\mu}R^{\rho}_{\nu}+\frac{2}{3}R_{\mu\nu}R
\right)_0\nonumber \\
&=& 2(\nabla_{A'}^C\nabla_{B'}^D+\Phi^{CD}_{A'B'} )\Psi_{ABCD}=2(\nabla_{A}^{C'}\nabla_{B}^{D'}+\Phi^{C'D'}_{AB})\widetilde{\Psi}_{A'B'C'D'},
\end{eqnarray}
where the subscript in the second line denotes `trace-free part.'  These show that the field equations are satisfied when $M$ is conformal to Einstein (i.e., $g_{\mu\nu}\propto R_{\mu\nu}$), or when its Weyl curvature is either self-dual or anti-self-dual.  

For Yang-Mills theory, there are Lagrangians which allow for a direct perturbative expansion around the self-dual sector based on a Lagrange multiplier action for the self-dual sector itself (c.f., \cite{Capovilla:1991qb, Chalmers:1996rq}).  Since the field equations for conformal gravity can be understood as the Yang-Mills equations of the Cartan conformal connection on a $\SU(2,2)$ bundle \cite{Merkulov:1984nz}, it is natural to expect that analogous actions exist for conformal gravity.  
We first note that we can add to \eqref{CGA1} the topological term 
\begin{equation*}
\frac{1}{\varepsilon^2}\int_{M}\d\mu\left(\Psi^{ABCD}\Psi_{ABCD}-\widetilde{\Psi}^{A'B'C'D'}\widetilde{\Psi}_{A'B'C'D'}\right) = \frac{12\pi^2}{\varepsilon^{2}}(\tau(M)-\eta(\partial M))\, ,
\end{equation*}
where $\tau(M)$ is the signature of $M$ and $\eta(\partial M)$ is the $\eta$-invariant of the conformal boundary \cite{Hitchin:1997}.  This gives the complex chiral action 
\be{CGA2}
S^{\mathrm{CG}}[g]=\frac{2}{\varepsilon^{2}}\int_{M}\d\mu\;\Psi^{ABCD}\Psi_{ABCD}\, ,
\ee
which is equivalent to the full action \eqref{CGA1} up to terms which are irrelevant in perturbation theory.  

To expand around the SD sector, we introduce the totally symmetric Lagrange multiplier spinor field $G_{ABCD}$ and write the action as \cite{Berkovits:2004jj}:
\be{CGA3}
S^{\mathrm{CG}}[g,G]=\int_{M}\d\mu \left(G^{ABCD}\Psi_{ABCD}-\varepsilon^{2}G^{ABCD}G_{ABCD}\right).
\ee 
This has field equations \cite{Mason:2005zm}
\be{CGFE}
\Psi^{ABCD}=\varepsilon^{2}G^{ABCD}, \qquad \left(\nabla^{C}_{A'}\nabla^{D}_{B'}+\Phi^{CD}_{A'B'}\right)G_{ABCD}=0,
\ee
so integrating out $G$ returns \eqref{CGA2}.  But now $\varepsilon^{2}$ becomes a parameter for expanding about the SD sector: when $\varepsilon=0$, the field equations yield a SD solution and $G_{ABCD}$ is a linear ASD solution propagating on the SD background.



\subsection{Einstein gravity amplitudes inside the conformal gravity S-matrix}

We now review  the relationship between conformal gravity and Einstein gravity and how it is manifested for scattering amplitudes on de Sitter backgrounds following an argument due to Maldacena \cite{Maldacena:2011mk}.  Similar arguments hold for anti-de Sitter space with some sign changes and in that form these ideas can be applied to AdS/CFT duality.  We review the geometry of de Sitter space in Appendix \ref{desitter-geom}.

It is easily seen from the definition of the Bach tensor that Einstein gravity solutions are also solutions to conformal gravity.  However, in order to show that Einstein tree amplitudes can be obtained from those of conformal gravity we need to relate the actions of the two theories.  This is because we can define the tree-level S-matrix (or at least its phase) to be the value of the action evaluated on a classical solution to the equations of motion that has been obtained perturbatively from the given fields involved in the scattering process.  More formally, given $n$ solutions $g_i$, $i=1,\ldots, n$ to the linearized field equations and a classical background $g^{\mathrm{cl}}$, we construct the solution $g$ to the field equations whose asymptotic data is $\sum_i \epsilon_i g_i$.  We can then define the tree amplitude to be
$$
\cM(1,\ldots, n)=\mbox { coefficient of } \prod_{i=1}^{n} \epsilon_i \mbox{ in } S[g^{\mathrm{cl}}+g]\, .
$$
Thus, if the conformal gravity action of a solution to the Einstein equations yields the Einstein-Hilbert action of that same solution, then the tree-level conformal gravity S-matrix can be used to compute that for general relativity.  We will see that this is the case up to a factor of $\Lambda$.

The Einstein-Hilbert action in the presence of a cosmological constant is
\begin{equation*}
S^{\mathrm{EH}}[g]=\frac{1}{\kappa^{2}}\int_{M}\d\mu (R-2\Lambda),
\end{equation*}
where $\kappa^{2}=16\pi G_{N}$.  On a de Sitter space, the field equations are $R_{\mu\nu}=\Lambda g_{\mu\nu}$, so the action reads
\begin{equation*}
S^{\mathrm{EH}}[dS_{4}]=\frac{2\Lambda}{\kappa^2}\int_{dS_{4}}\d\mu =\frac{2\Lambda}{\kappa^2}V(dS_{4}),
\end{equation*}
where $V(M)$ is the volume of $M$.  For any asymptotically de Sitter manifold, this volume will be infinite so the action functional must be modified by the Gibbons-Hawking boundary term \cite{Gibbons:1976ue}.  Additionally, we must include the holographic renormalization counter-terms (which also live on the boundary) in order to render the volume finite \cite{Balasubramanian:1999re, Skenderis:2002wp}.  After including these additions, one obtains the \emph{renormalized} Einstein-Hilbert action \cite{Miskovic:2009bm}, and if $M$ is asymptotically de Sitter, we have:
\be{EHvol}
S^{\mathrm{EH}}_{\mathrm{ren}}[M]=\frac{2\Lambda}{\kappa^2}V_{\mathrm{ren}}(M),
\ee
where $V_{\mathrm{ren}}$ is the renormalized volume of the space-time (c.f., \cite{Graham:1999jg}).  

On the other hand, if $M$ was a Riemannian 4-manifold which was compact without boundary, the Chern-Gauss-Bonnet formula would tell us that
\begin{equation*}
\chi(M)=\frac{1}{8\pi^{2}}\int_{M}\d\mu \left(C^{\mu\nu\rho\sigma}C_{\mu\nu\rho\sigma}-\frac{1}{2}R_{\mu\nu}R^{\mu\nu}+\frac{1}{6}R^{2}\right).
\end{equation*}
If $M$ were additionally Einstein $(R_{\mu\nu}=\Lambda g_{\mu\nu})$, then we would have
\be{CGB1}
S^{\mathrm{CG}}[M]=\frac{8\pi^{2}\chi(M)}{\varepsilon^{2}}-\frac{2\Lambda^{2}}{3\varepsilon^{2}}V(M).
\ee
When $M$ is (Lorentzian) asymptotically de Sitter, the Chern-Gauss-Bonnet formula requires a boundary term, and the volume is renormalized.  However, a theorem of Anderson\footnote{Note that Anderson's theorem is actually stated for asymptotically hyperbolic Riemannian four-manifolds; the extension to asymptotically de Sitter Lorentzian manifolds follows by analytic continuation.} tells us that \eqref{CGB1} continues to hold even after boundary terms for the Euler characteristic are taken into account and the volume has been renormalized \cite{Anderson:2001}.  Furthermore, since $M$ is asymptotically de Sitter we can assume that we always perturb around the topologically trivial case (i.e., $\chi(M)=0$), so comparing with \eqref{EHvol} we find
\be{CGB2}
S^{\mathrm{CG}}[M]=-\frac{\Lambda\;\kappa^{2}}{3\;\varepsilon^{2}}S^{\mathrm{EH}}_{\mathrm{ren}}[M].
\ee

In Appendix \ref{GenFuncs}, we show that this embedding can be made precise at the level of generating functionals for the MHV amplitudes of the two theories.  In particular, proposition \ref{CGDS} demonstrates that the Einstein generating functional is equivalent to the one from conformal gravity, up to the constants predicted by \eqref{CGB2} and restriction to Einstein degrees of freedom.  Additionally, these ideas extend to a particular phenotype of $\cN=4$ conformal supergravity (CSG), known as the \emph{minimal} theory.  This version of $\cN=4$ CSG posesses a global $\SU(1,1)$ symmetry acting non-linearly the conformal dilaton; this excludes any graviton-scalar couplings which have no analogue in Einstein gravity.  Appendix \ref{minimal} provides a review of these concepts.


\subsection{Relations between Einstein and conformal gravity polarization states}
\label{PolarStates}

Maldacena also argues that we can single out Einstein scattering states inside conformal gravity by employing boundary conditions on the metric \cite{Maldacena:2011mk}.  We will use an equivalent explicit formulation in twistor space to compute the tree-level scattering amplitudes of general relativity by using conformal gravity restricted to Einstein scattering states on a de Sitter background.  This is realized on space-time as follows.

The usual strategy for calculating scattering amplitudes is to express them in terms of a basis of momentum polarization states.  We will in fact use a variety of different representations arising from twistor space; however, we need some understanding of the relationship between linearized solutions to the Bach equations \eqref{Bach}, spin-two fields, and linearized Einstein solutions.  Polarization states for conformal gravity were studied in \cite{Berkovits:2004jj, Mannheim:2009qi} and were argued to contain twice as many states as for Einstein gravity.  The representation on twistor space shows that there are actually three times as many conformal gravity states as for Einstein gravity.  In Appendix \ref{pol-states} we give a momentum space argument for this; presumably one has simply been missed in earlier treatments.  This will not materially alter our discussion here, as we are concerned more with the embedding of Einstein states into those for conformal gravity, though.  

We use a slightly different formulation from previous treatments that allows us to retain Lorentz invariance (although not translation invariance), and will also tie in with our focus on de Sitter gravity.   Let $\{\psi_{ABCD}, \tilde{\psi}_{A'B'C'D'}\}$ be linearized spin-two fields and $\{\Psi_{ABCD}, \tilde{\Psi}_{A'B'C'D'}\}$ be the ASD and SD portions of the Weyl tensor.  The key point in connecting conformal gravity to spin-two fields is that the Weyl tensor has conformal weight zero, whereas a linearized spin-two field has conformal weight $-1$ (c.f., \cite{Penrose:1986ca}).  Both fields satisfy
\be{spin-two}
\nabla_A^{A'}\tilde{\Psi}_{A'B'C'D'}=\nabla_A^{A'}\tilde{\psi}_{A'B'C'D'}=0=\nabla_{A'}^A\Psi_{ABCD}=\nabla_{A'}^A\psi_{ABCD}\, ,
\ee 
in the Einstein conformal frame but the Weyl tensor only does so in its given Einstein conformal scale and no other.  Einstein conformal scales can be specified as functions $\Omega$ of conformal weight $+1$ that satisfy the conformally invariant equation \cite{Lebrun:1985}
\be{E-scale}
(\nabla_{\mu}\nabla_{\nu} + \Phi_{\mu\nu})_0\Omega=0,
\ee
where the subscript $0$ denotes `the trace-free part' and $\Phi_{\mu\nu}$ is half the trace-free part of the Ricci tensor.

In flat space, \eqref{E-scale} has the general solution 
\be{Einstein-scale-soln}
\Omega=a+b_{\mu} x^{\mu}+ c x^2\, .
\ee
It is clear in general that given such a solution $\Omega$, rescaling so that $\Omega=1$
gives a metric satisfying $\Phi_{\mu\nu}=0$ from \eqref{E-scale}.
This is the Einstein condition, and the solutions \eqref{Einstein-scale-soln} give metrics with cosmological constant $\Lambda=3(b_\mu b^\mu- ac)$.
Upon setting 
\be{spin-two-E}
\Psi_{ABCD}=\Omega\psi_{ABCD} ,
\ee
we see that the Weyl spinor $\Psi_{ABCD}$ has conformal weight zero and satisfies the linearized vacuum Bianchi identity \eqref{spin-two} for the conformal scale in which $\Omega=1$.  Since this is an Einstein scale and the Bach equations are simply another derivative of this equation, $\Psi_{ABCD}$ also satisfies the linearized Bach equations.  But then, by conformal invariance of the Bach equations, it does so in \emph{any} conformal scale.

We refer the reader to Appendix \ref{pol-states} for some further discussion of momentum eigenstates that will not be needed in what follows. 

\section{Twistor Action for Conformal (Super-)Gravity}
\label{TAM}

In this section, we show how conformal gravity and its supersymmetric extension can be formulated in terms of a classical action functional on twistor space.  After first recalling some background material on twistor spaces for curved space-times, we define the twistor action for $\cN=0$ conformal gravity \cite{Mason:2005zm} and then consider its natural extension to $\cN=4$ supersymmetry.  Our treatment here is rather different from that in \cite{Mason:2005zm} as we focus on a coordinate description that has a simple perturbative expansion.


\subsection{Curved twistor theory}

In flat Minkowski space $\M$, twistor space $\PT$ is an open subset of $\CP^3$, with homogeneous coordinates $Z^{\alpha}=(\lambda_{A},\mu^{A'})$.  The standard flat-space incidence relations
\begin{equation*}
\mu^{A'}=ix^{AA'}\lambda_{A},
\end{equation*}
represent a point $x\in\M$ by a linearly embedded $\CP^{1}\subset\PT$.  To study conformal gravity and the MHV generating functional \eqref{CGDS*}, we need twistor theory adapted to curved space-times such as the self-dual background with cosmological constant, $M$.   

The \emph{non-linear graviton} construction is the basis for curved twistor theory.  We state the theorem in the context of $\cN=0$, but its extension to the $\cN=4$ context is straightforward.
\begin{thm}[Penrose \cite{Penrose:1976js}, Ward \cite{Ward:1980am}]\label{NLG}
There is a one-to-one correspondence between: \emph{(a.)} Space-times
$M$ with self-dual conformal structure $[g]$, and  \emph{(b.)} twistor
spaces $\CPT$ (a complex projective 3-manifold) obtained as a complex deformation of $\PT$ and containing at least one rational curve $X_0$ with normal bundle $N_{X_0}\cong\cO(1)\oplus \cO(1)$.  Define the complex line bundle $\cO(1)\rightarrow\CPT$ so that $\Omega^{3}_{\CPT}\cong\cO(-4)$ (the appropriate $4^{\mathrm{th}}$ root exists on the neighbourhood of $X_0$ from the previous assumption). 

There is a metric $g\in [g]$ with Ricci curvature $R_{\mu\nu}=\Lambda g_{\mu\nu}$ if and only if $\CPT$ is equipped with:  
\begin{itemize}
\item a non-degenerate holomorphic contact structure specified by $\tau\in\Omega^{1,0}_{\CPT}(2)$, and
\item a holomorphic 3-form $\D^{3}Z\in\Omega^{3,0}_{\CPT}(4)$ obeying $\tau\wedge \d\tau=\frac{\Lambda}{3}\D^{3}Z$.
\end{itemize}
Here $\D^{3}Z$ is the tautologically defined section of $\Omega^3_{\CPT}(4)$.

We define the non-projective twistor space $\CT$ to be the total space of the complex line bundle $\cO(-1)$.
\end{thm}

Thus, points $x\in M$ (for $M$ obeying the conditions of this theorem)
correspond to rational, but no longer necessarily linearly embedded, curves
$X\subset\CPT$ of degree 1.  The conformal structure on $M$ corresponds to
requiring that if two of these curves $X$, $Y$ intersect in $\CPT$,
then the points $x,y\in M$ are null separated.  Furthermore, $\CPT$ can be reconstructed as the space of totally null self-dual 2-planes in the
complexification of $M$ (c.f., \cite{Penrose:1976js, LeBrun:1982}).  

Theorem \ref{NLG} tells us that $M$ corresponds to a curved twistor space $\CPT$ which arises as a complex deformation of $\PT$.  We will take $M$ to be a finite but small perturbation away from flat space, so the deformed complex structure on $\CPT$ will be  expressed as a small but finite deformation of the flat $\dbar$-operator:
\begin{equation*}
\dbar_{f}=\dbar +f= \d \bar{Z}^{\bar{\alpha}} \frac{\partial}{\partial \bar{Z}^{\bar{\alpha}}}+ f,
\end{equation*}
where $f\in\Omega^{0,1}_{\PT}(T_{\PT})$ and $Z^{\alpha}$ are homogeneous coordinates on $\CPT$.  This induces a basis for $T^{0,1}_{\CPT}$ and $\Omega^{1,0}_{\CPT}$ with respect to the deformed complex structure:
\begin{eqnarray}
T^{0,1}_{\CPT}=\mathrm{span}\left\{\frac{\partial}{\partial\bar{Z}^{\bar{\alpha}}}+f^{\alpha}_{\bar{\alpha}}\frac{\partial}{\partial Z^{\alpha}}\right\}, \label{gbasis} \\
\Omega^{1,0}_{\CPT}=\mathrm{span}\{\D Z^{\alpha}\}=\mathrm{span}\left\{\d Z^{\alpha}-f^{\alpha}\right\}, \label{gfbasis}
\end{eqnarray}
where we have denoted $f=f^{\alpha}\partial_{\alpha}=f^{\alpha}_{\bar{\alpha}}\d \bar{Z}^{\bar{\alpha}}\partial_{\alpha}$.  The forms $f^{\alpha}$ must descend from $\mathscr{T}$ to $\CPT$, which follows from 
\be{fgf}
\bar{Z}^{\bar{\alpha}}f^{\beta}_{\bar{\alpha}}=0\, , \qquad f^\alpha(\lambda Z)=\lambda f^\alpha(Z)\,,\quad \lambda\in\C^{*}.
\ee 
Additionally, the vector field $f$ on $\CT$ is determined by one on $\CPT$ only up to multiples of the Euler vector field $Z^{\alpha}\partial_{\alpha}$, and this freedom can be fixed by imposing
\be{fgf1}
\partial_{\alpha}f^{\alpha}=0\, .
\ee

As it stands, $\dbar_f$ defines an almost complex structure.  This is integrable if and only if
\be{integrability}
\dbar f^{\alpha}+\left[f,f\right]^{\alpha}=0, \qquad \left[f,f\right]^{\alpha}=f^{\beta}\wedge\partial_{\beta}f^{\alpha}\, .
\ee
This integrability condition can be thought of as the twistor form of the field equations for self-dual conformal gravity.  Kodaira theory implies the existence of a complex four parameter family of rational curves of degree one, and this family is identified with the complexification of space-time $M$.  Thus to reconstruct $M$ from $\CPT$ we must find a family of holomorphic maps
\begin{equation*}
Z^{\alpha}(x^{\mu},\sigma_{A}):\PS\rightarrow \CPT, \qquad Z^{\alpha}(x,\sigma)=\left(\lambda_{A}(x,\sigma), \mu^{A'}(x,\sigma)\right), 
\end{equation*}
where $\PS\cong M\times\CP^{1}$ is naturally identified with the un-primed projective spinor bundle of $M$ and $Z(x,\sigma)$ is a map of degree one parametrized by $x\in M$.  We will often denote the image of the map for $x\in M$ as $X$.  The condition that these maps be holomorphic is
\be{holomap}
\dbar_{\sigma} Z^{\alpha}(x,\sigma)-f^{\alpha}(Z(x,\sigma))=0,
\ee
where $\dbar_{\sigma} =\d\bar{\sigma}\frac{\partial}{\partial\bar{\sigma}}$ is the $\dbar$-operator on $X\subset\CPT$ pulled back to $\PS$.  


\subsection{Twistor action}

We construct a twistorial version of the chiral action \eqref{CGA3} in twistor space in two parts.  The first is an action for the self-dual sector of conformal gravity.  By theorem \ref{NLG}, this is equivalent to a twistor space with almost complex structure $\dbar_{f}$ subject to the field equation that it be integrable.  The integrability condition is the vanishing of
\be{Nijenhuis}
\dbar_{f}^{2} = \left(\dbar f^{\alpha}+[f,f]^{\alpha}\right)\partial_{\alpha}\in\Omega^{0,2}_{\CPT}(T_{\CPT})\, .
\ee
This will follow as the field equations from the Lagrange multiplier action \cite{Berkovits:2004jj}:
\be{SDTA}
S_{1}[g,f]=\int_{\CPT}\D^{3}Z\wedge g_{\alpha}\wedge\left( \dbar f^{\alpha}+\left[f,f\right]^{\alpha}\right),
\ee
where $g:=g_{\alpha}\D Z^\alpha \in\Omega^{0,1}_{\CPT}(\cO(-4)\otimes \Omega^1)$ and is subject to $Z^{\alpha}g_{\alpha}=0$ because $f^\alpha$ is defined modulo $Z^\alpha$.\footnote{If we fix this freedom in $f^\alpha$ so that $\p_\alpha f^\alpha=0$, then we can allow a gauge freedom $g_\alpha \rightarrow g_\alpha + \p_\alpha \chi$, although this makes less geometric sense as then $g$ becomes non-projective.}  The field equations for this action are
\be{SDFEs}
\dbar f^{\alpha}+\left[f,f\right]^{\alpha}=0, \qquad \dbar_{f}\left(g_{\alpha}\D Z^{\alpha}\right)=0\, .
\ee

We additionally have the gauge freedom $g\rightarrow g + \dbar_f \alpha$ for $\alpha\in \Omega^1_{\CPT}(-4)$ due to a Jacobi-like identity for the almost complex structure.  Thus, on-shell at least,  $g$ defines a cohomology class in $H^{0,1}(\CPT, \Omega^1(-4))$.  We can therefore apply the Penrose transform \cite{Hitchin:1980hp} to define a space-time field $G_{ABCD}$ by:
\be{CSF1}
G_{ABCD}(x)=\int_{X} \lambda_{A}\lambda_{B}\lambda_{C}\lambda_{D}\;g(Z(x,\sigma)).
\ee
It is straightforward to show that $G_{ABCD}$ satisfies the second field equation of \eqref{CGFE} using the properties of the Penrose transform (c.f., \cite{Adamo:2013}).  Thus $g$ gives rise to a linear ASD conformal gravity field propagating on the SD background.  

The action \eqref{SDTA} is therefore equivalent to the first (self-dual) term of the chiral space-time action \eqref{CGA3}, i.e.,  with $\varepsilon^{2}=0$.  To obtain the ASD interactions of the theory, we simply need to express the second term in \eqref{CGA3} in twistor space.  The Penrose transform \eqref{CSF1} can be implemented off-shell to give:
\be{TAInt}
S_{2}[g,f]=\int_{\PS\times_{M}\PS}\d^4x \wedge \la \lambda_{1}\;\lambda_{2}\ra^{4}\;g_{1}\wedge g_{2},
\ee
where $\PS\times_{M}\PS\cong M\times \CP^1\times \CP^1$ is the fibre-wise product of $\PS$ with itself, and $\rd^4x$ is an integration measure that, as we will see later, is canonically defined.  In this expression for $S_2$, we implicitly assume that the SD background $M$ is constructed via the non-linear graviton of theorem \ref{NLG}.  This can be made explicit by introducing a Lagrange multiplier field $Y\in\Omega^{1,0}_{\CP^1}(T^{*}\CPT)$ and re-writing the action as
\be{TAInt2}
S_{2}[g,f]=\int_{M}\d^4x \left[\int_{X} Y_\alpha \left(\dbar_{\sigma} Z^{\alpha}-f^{\alpha}\right)+\int_{X\times X}\la \lambda_{1}\;\lambda_{2}\ra^{4}\;g_{1}\wedge g_{2}\right].
\ee
Integrating out the field $Y_{\alpha}$ produces the constraint $\dbar_{\sigma} Z^{\alpha}=f^{\alpha}$, matching \eqref{holomap} and returning \eqref{TAInt}.  Note that the Lagrange multiplier $Y$ appears in a similar fashion in the worldsheet action of the Berkovits-Witten twistor-string \cite{Berkovits:2004hg, Berkovits:2004jj}.  

This gives the twistor action for the full (i.e., non-self-dual) conformal gravity of the form:
\be{CGTA}
S[g,f]=S_{1}[g,f]-\varepsilon^{2}S_{2}[g,f].
\ee
We should note that to define the action off shell, we must nevertheless solve \eqref{holomap} in order to define the integrals in $S_2$.  This equation can be solved with the standard four complex dimensional family of solutions irrespective of whether the almost complex structure is integrable \cite{Penrose:1976js, Hansen:1978jz}.  However, the integral against $\d^{4}x$ in \eqref{TAInt} is over a \emph{real} four-dimensional contour, so we must also impose a reality condition on the data in order for the moduli space of solutions to have a real four-dimensional slice.  This can be done by imposing a reality structure that is adapted to either Euclidean or split signature.  For Euclidean signature we have an anti-linear involution $Z^\alpha\rightarrow \hat Z^\alpha$ that is quaternionic so that $\hat{\hat Z}^\alpha=-Z^\alpha$ and we require $\bar{f}= f(\hat Z)$.  This induces a conjugation on $M$ whose fixed points are a real slice of Euclidean signature (an ordinary conjugation yields a real slice of split signature).

The following theorem confirms that \eqref{CGTA} is equivalent to \eqref{CGA3}, as desired:
\begin{thm}[Mason \cite{Mason:2005zm}]\label{MThm}
The twistor action $S[g,f]$ is classically equivalent to the conformal gravity action \eqref{CGA3} off-shell  in the sense that there exists a gauge in which it reduces to the space-time action.  In particular, solutions to its Euler-Lagrange equations are in one-to-one correspondence with solutions to the field equations \eqref{CGFE} up to space-time diffeomorphisms and it correctly gives the value of the action evaluated on such solutions.
\end{thm}

We refer to \cite{Mason:2005zm} for the proof.  While we have expressed \eqref{CGTA} with respect to a choice of background complex structure (for comparison to the Einstein case below), the conformal gravity twistor action can be formulated in a coordinate-invariant way.  The theorem is then proven by going to a coordinate system $(x^{\mu},\sigma_{A})$ in which \eqref{holomap} is solved for the pseudo-holomorphic curves.  In that gauge, fields can be integrated out or compared directly to their space-time counterparts.


\subsection{The $\cN=4$ minimal twistor action}

The extension of the above construction to $\cN=4$ supersymmetry is straightforward.  The twistor space $\CPT$ becomes a projective $(3|4)$-dimensional supermanifold modeled on $\CP^{3|4}$ with homogeneous coordinates $Z^I=(Z^\alpha, \chi^a)$, $a=1,\ldots , 4$.  It is super-Calabi-Yau being equipped with a (canonical) holomorphic volume measure $\D^{3|4}Z$ (i.e., a canonical holomorphic section of the Berezinian).  The data naturally extends to a deformed $\bar\p$-operator and $(1,1)$-form on $\CPT$ 
$$
\dbar_f=\dbar+ f^I\frac\p{\p Z^I} \, , \qquad g:=g_I \D Z^I\in \Omega^{1,1}_{\CPT}\, , \quad \D Z^I=\rd Z^I-f^I\, .
$$
With $\cN=4$ supersymmetry, the conditions $\partial_{I}f^{I}=0$ and $Z^{I}g_{I}=0$ no longer fix the gauge freedoms of adding a multiple of $Z^I$ to $f^I$ or $\p_{I}$ to $g$.  Since $\partial_{I}f^{I}=0$ on account of fermionic signs, $\partial_{I}f^{I}=0$ is compatible with adding a multiple of $Z^{I}$ to $f^{I}$, and $Z^{I}g_{I}=0$ is compatible with adding $\partial_{I} \alpha$ to $g$, as $\alpha$ now has homogeneity zero rather than $-4$.   

This allows us to define \eqref{CGTA} with respect to the new super-geometry by taking:
\be{minTA1}
S_{1}[g,f]=\int_{\CPT}\D^{3|4}Z\wedge g_{I}\wedge \left(\dbar f^{I}+[f,f]^{I}\right),
\ee
\be{minTA2}
S_{2}[g,f]=\int\limits_{\PS\times_{M}\PS}\d ^{4|8}x\wedge g_{1}\wedge g_{2}.
\ee
Again, as we will see in section \ref{MEAS}, $\d^{4|8}x$ is a canonically defined measure on the $(4|8)$-dimensional chiral space-time $M$, the space of degree-one rational curves in $\CPT$.  When restricting to the degrees of freedom of bosonic conformal gravity, the fermionic integrals just have the effect of producing the factor of $\la \lambda_1\lambda_2\ra^4$ in \eqref{TAInt}.  As in the $\cN=0$ setting, we can make the construction of the SD background $M$ explicit by introducing the Lagrange multiplier $Y$ and writing
\be{minTA3}
S_{2}[g,f]=\int_{M}\d^{4|8}x \left[\int_{X}Y_I\left(\dbar_{\sigma}Z^{I}-f^{I}\right)+\left(\int_{X}g\right)^2\right].
\ee

In the supersymmetric setting, $g_{I}\D Z^{I}$ defines a chiral superfield on space-time:
\be{s-tfield}
\mathcal{G}(x,\theta)=\int_{X} g(Z(x,\theta,\sigma)),
\ee
where $\mathcal{G}$ has an expansion like:
\begin{equation*}
\mathcal{G}(x,\theta)=\varphi +\cdots +\theta^{4\;ABCD}\Psi_{ABCD}+\cdots .
\end{equation*}
The Penrose transform can be used to show that the individual fields in $\mathcal{G}$ correspond to the chiral (ASD) half of the $\cN=4$ CSG field content, as desired.  The space-time translation of our $\cN=4$ twistor action will look like
\be{CSUGRAct}
S[\mathcal{W},\mathcal{G}]=\int_{M}\d ^{4|8} x \left(\mathcal{W}(x,\theta)\;\mathcal{G}(x,\theta)-\varepsilon^{2}\mathcal{G}(x,\theta)^{2}\right) \rightarrow \frac{1}{\varepsilon^2}\int_{M}\d^{4|8}x \; \mathcal{W}(x,\theta)^2,
\ee
where $\mathcal{W}(x,\theta)$ is the a chiral superfield which, on-shell, is a Lorentz scalar encoding the $\cN=4$ Weyl multiplet (c.f., \cite{Bergshoeff:1980is}).  

This action has the correct linear reduction for $\cN=4$ CSG \cite{Berkovits:2004jj}, and must correspond to a \emph{minimal} theory since the functional form prohibits any cubic couplings between $\varphi$ and the Weyl curvature (see Appendix \ref{minimal} for a discussion).  However, note that our twistor action only possesses the linearized $\E_{2}$ global symmetry of translating the scalar $\varphi$ rather than the fully non-linear $\SU(1,1)$ of a generic minimal theory.\footnote{The additional $\U(1)$-symmetry of the minimal model can be seen as arising from $g\rightarrow\e^{4i\beta}g$ together with $\chi^{a}\rightarrow\e^{-i\beta}\chi^{a}$ which induces a similar phase rotation for $\theta^{aA}$.  This symmetry is the key for ruling out the $\varphi\; (\mathrm{Weyl})^2$ couplings and hence ensuring that the embedding of Einstein gravity still applies.}  Nevertheless, Einstein supergravity still forms a subsector of this degenerate theory \cite{Cremmer:1977tt}, so the embedding of Einstein states still applies.  


\subsection{The volume form}
\label{MEAS}

To complete the definition of the twistor action, we must specify the volume form $\rd^{4|8}x$ used in \eqref{minTA2}; its reduction to $\rd^4x $ in the non supersymmetric case will follow directly from this.  To this end we rewrite \eqref{holomap} as an integral equation
\be{Picard1}
Z^{I}(x,\sigma)=X^{I}_{A}\sigma^{A}+\dbar_{\sigma}^{-1}\left(f^{I}(Z)\right)\, ,
\ee
where $X^{I}_{A}\sigma^{A}$ solves the homogeneous equation and $X^{IA}$ parametrizes its solutions.  Since $f^{I}$ has weight $+1$, there is an ambiguity in the choice of $\dbar_{\sigma}^{-1}$, which can be chosen to vanish at two points on $\CP^1$.  For simplicity we will require that it vanishes at $\sigma_A=\xi_A$ to second order by setting 
\be{Picard2}
Z^{I}(x,\sigma)=X^{I}_{A}\sigma^{A}+\frac1{2\pi i}\int_{\CP^1}\frac{\D\sigma'}{(\sigma\sigma')}\frac{(\xi\sigma)^2}{(\xi\sigma')^2}f^{I}(Z(\sigma')).
\ee
Physical observables such as scattering amplitudes will be independent of $\xi$ at the end of our calculations and we will perform this consistency check explicitly.

We now write $Z^{I}(x,\sigma)=\mathcal{X}^{I}_{A}\sigma^{A}$ defining
\be{Picard3}
\mathcal{X}^{I A}(x,\sigma)=X^{I A}+\frac{\xi^{A}}{2\pi i}\int_{\CP^1}\frac{\D\sigma'}{(\sigma\sigma')}\frac{(\xi\sigma)}{(\xi\sigma')^2}f^{I}(Z(\sigma'))\, ,
\ee
which solves 
\be{Picard4}
\dbar_{\sigma} \mathcal{X}^{I A}(x,\sigma)=\frac{\xi^{A}f^{I}}{(\xi\sigma)}.
\ee
This enables us to take the exterior derivative of $\mathcal{X}$ with respect to the space-time coordinate $x$, finding
\be{Yder}
\dbar_{\sigma}\left(\d_{x}\mathcal{X}^{I A}(x,\sigma)\right)=\partial_{J} f^{I}\frac{\xi^{A}\sigma_{B}}{(\xi\sigma)}\d_{x}\mathcal{X}^{J B}(x,\sigma).
\ee
Since $\partial_{I}f^{I}=0$, this means that the top-degree form $\d^{8|8}\mathcal{X}$ is holomorphic in $\sigma$ and of weight zero; by Liouville's theorem, it is therefore independent of $\sigma$.  But this means that
\begin{equation*}
\d^{4|8}x\equiv\frac{\d^{8|8}\mathcal{X}}{\mathrm{vol}\;\GL(2,\C)}=\frac{\d^{8|8}X}{\mathrm{vol}\;\GL(2,\C)},
\end{equation*}
is an invariant volume form on the space-time $M$ itself.\footnote{Here $\GL(2,\C)$ is the choice of homogeneous coordinates $\sigma_A$ on $X\cong\CP^{1}$.  The division by $\mathrm{vol}\;\GL(2,\C)$ is understood in the Fadeev-Popov sense: one chooses a section of the group action, and multiplies by the appropriate Jacobian factor to obtain a well-defined volume form on the $(4|8)$-dimensional quotient.  One can also define this form to be that obtained by contracting a basis set of the generators of $\GL(2,\C)$ into the volume form in the numerator and observing that the form is one pulled-back from the quotient.}


\section{Reduction to Einstein Gravity}
\label{TAM2}

In this section, we reduce the degrees of freedom in the twistor action for conformal (super)gravity to those of Einstein gravity off-shell.  We then use the embedding of Einstein gravity into conformal gravity described earlier to argue that, upon division by $\Lambda$ we have the correct Einstein action.  For the sake of convenience, we work primarily in the $\cN=4$ formulation, although the translation to $\cN=0$ conformal gravity and general relativity should be obvious.  

We first illustrate how the twistor data for $\cN=4$ CSG is reduced to the Einstein subsector.  This produces the conformal gravity twistor action restricted to Einstein data, and in particular leads to a twistorial expression for the MHV generating functional $I^{\mathrm{CG}}$ derived in \eqref{CGGF}.  It also leads to a new action functional which we propose describes Einstein gravity itself.  This new twistor action has the correct self-dual sector, can be defined for $\cN=0,4,8$, and also leads to the correct expression for the MHV amplitude, as we demonstrate in Section \ref{MACC}.


\subsection{The Einstein degrees of freedom}

We now reduce the data of the $\cN=4$ CSG twistor action to the Einstein subsector.  This will be done off-shell in the first instance.  A conformal factor $\Omega$ from \eqref{Einstein-scale-soln} relating spin-two and linearized Einstein fields can be specified on twistor space by introducing an \emph{infinity twistor} $I_{IJ}$, a skew bi-twistor.\footnote{In the supersymmetric case, the fermionic part of the infinity twistor corresponds to a gauging of the $\cN=4$ $R$-symmetry \cite{Wolf:2007tx}; this will not play an important role in this paper.}  


Choose $I_{0}$ and $I_{1}$ to be of rank-two such that
\begin{equation*}
I_{0\;IJ}Z^{I}\d Z^{J}=\la \lambda\;\d\lambda\ra, \qquad I_{1\;IJ}Z^{I}\d Z^{J}=[\mu\;\d\mu].
\end{equation*}
Then the infinity twistor appropriate to Einstein polarization states with cosmological constant $\Lambda$ on the affine de Sitter patch is given by $I=I_{0}+\Lambda I_{1}$.   We can define an upstairs bosonic part by  $I^{\alpha\beta}=\frac{1}{2}\epsilon^{\alpha\beta\gamma\delta}I_{\gamma\delta}$ and we will have 
$$
I_{\alpha\beta}I^{\beta\gamma}=\Lambda \delta_\alpha^\gamma\, .
$$
This relation can be extended supersymmetrically if we set:
\be{inftyCC}
I_{IJ}=\left(
\begin{array}{ccc}
\epsilon^{AB} & 0 & 0 \\
0 & \Lambda \epsilon_{A'B'} & 0 \\
0 & 0 & \sqrt{\Lambda}\delta_{ab}
\end{array} \right), \qquad 
I^{IJ}=\left(
\begin{array}{ccc}
\Lambda \epsilon_{AB} & 0 & 0 \\
0 & \epsilon^{A'B'} & 0 \\
0 & 0 &  \sqrt{\Lambda}\delta^{ab}
\end{array}\right).
\ee  
For brevity we introduce the notation
\be{brackets}
[A, B]:= I^{IJ}A_IB_J\, , \qquad \la C, D\ra :=I_{IJ}C^ID^J\, .
\ee

Geometrically, these infinity twistors are encoded into a weighted contact form $\tau$ and Poisson structure on $\CPT$:
\be{infstruct}
\tau= I_{IJ}Z^{I}\D Z^{J}=\la Z,\D Z\ra \qquad
\Pi=I^{IJ}\partial_{I}\wedge\partial_{J}, \qquad
\{f,g\}=I^{IJ}\partial_{I}f\;\partial_{J}g=[\p f,\p g]\, .
\ee
Here, $\Pi$ is the Poisson bi-vector and $\{\cdot, \cdot\}$ is the corresponding Poisson bracket.

We now require that the complex structure $\dbar_{f}$ be Hamiltonian with respect to $\Pi$, $\cL_{f}\Pi =0$, and this implies that
\begin{equation*}
f^I=I^{IJ}\partial_{J}h, \qquad h\in\Omega^{0,1}_{\CPT}(2).
\end{equation*}
The integrability condition for such an almost complex structure is the vanishing of
\be{ham-int} 
\dbar_f^2=I^{IJ}\partial_{J}\left(\dbar h +\frac{1}{2} \left\{h,h\right\}\right)\partial_{I}\, .
\ee
The remaining diffeomorphism freedom on $\CPT$ is captured by the infinitesimal transformations:
\begin{equation*}
\delta Z^{\alpha}=\left\{Z^{\alpha},\chi\right\}, \qquad \delta h=\dbar\chi +\left\{h,\chi\right\},
\end{equation*}
for $\chi$ a weight $+2$ function \cite{Mason:2007ct}.

In the linearized setting, we have $\dbar h=0$ and $h$ is defined modulo infinitesimal Hamiltonian diffeomorphisms, so $h$ defines a cohomology class in $H^{0,1}(\CPT,\cO(2))$.  The Penrose transform realizes this as a $\cN=4$ graviton multiplet of helicity $+2$ via the integral formula
$$
\tilde{\psi}(x,\theta)_{A'B'C'D'}=\int_X \frac{\p^4h}{\p\mu^{A'}\cdots \p\mu^{D'}}\wedge\tau\, .
$$
Dually, we have the relation $g_I =I_{IJ}Z^J\tilde h$ with $\tilde h\in \Omega^{0,1}_{\CPT}(-2)$.   In the linearized theory, $\tilde{h}$ will define a cohomology class in $H^{0,1}(\CPT,\cO(-2))$ and the Penrose transform identifies this with the on-shell $\cN=4$ graviton multiplet of helicity $-2$ \cite{Adamo:2013}, this time starting with the scalar
$$
\phi(x,\theta)=\int_X \tilde{h}\wedge \tau \, .
$$  


\subsection{The action and MHV generating functional}

We can now insert the off-shell Einstein data  $f^I=I^{IJ}\p_J h$  and $g_I=I_{IJ}Z^J\tilde h$ into the $\cN=4$ CSG twistor action.  In the self-dual part of the action  \eqref{minTA1} we get:
\begin{multline}\label{EinR1}
S_{1}[g,f]\rightarrow S_{1}[g,h]=\int_{\CPT}\D^{3|4}Z\wedge g_{I}\wedge I^{IJ}\partial_{J}\left(\dbar h +\frac{1}{2} \left\{h,h\right\}\right) \\
=\int_{\CPT}\D^{3|4}Z\wedge I^{IJ}\partial_{I}g_{J}\wedge\left(\dbar h +\frac{1}{2} \left\{h,h\right\}\right) = 2\Lambda \int_{\CPT}\D^{3|4}Z\wedge\tilde{h}\wedge\left(\dbar h +\frac{1}{2} \left\{h,h\right\}\right)\, ,
\end{multline}
with the second line following via integration by parts.  This is precisely the self-dual twistor action for Einstein gravity, up to the factor of $\Lambda$ required by the embedding of Einstein gravity into conformal gravity \cite{Mason:2007ct, Adamo:2012nn}.  

The Einstein reduction for the second term of the twistor action \eqref{minTA2} is simply:
\begin{equation}\label{EinR3}
S_{2}[g,f]\rightarrow S_{2}[\tilde{h},h]
=\int_{M}\d^{4|8}x\left[\int_{X}Y_{I}\left(\dbar_{\sigma}Z^{I}-I^{IJ}\partial_{J}h\right)+\varepsilon^{2}\left(\int_{X} \tilde{h}\wedge\tau\right)^2\right],
\end{equation}
giving the explicit construction of $M$ via the non-linear graviton.  From our discussion in Appendix \ref{GenFuncs}, we know that this is the MHV generating functional for conformal gravity, restricted to Einstein states.  Then by proposition \ref{CGDS}, it follows that this should provide the generating functional for \emph{Einstein gravity} MHV amplitudes:
\begin{equation*}
I^{\mathrm{GR}}[1^{-},2^{-},M^{+}]=-\frac{3\varepsilon^{2}}{\Lambda\;\kappa^{2}} S_{2}[\tilde{h},h].
\end{equation*}
The perturbative content of the $n-2$ positive helicity gravitons is encoded by $h$, and in the next section we describe in detail the Feynman diagram calculus which allows us to recover the $n$-point MHV amplitude from this expression.
  
The restriction to Einstein states does more than provide us with a twistorial expression for the MHV generating functional.  Combining \eqref{EinR1} and \eqref{EinR3}, we can divide by a power of $\Lambda$ in accordance with the embedding of Einstein gravity into conformal gravity  to define an \emph{Einstein twistor action} (for $\cN=4$):
\begin{multline}\label{EinTA4}
S^{\mathrm{Ein}}_{\cN=4}[\tilde{h},h]=\int_{\CPT}\D^{3|4}Z\wedge\tilde{h}\wedge\left(\dbar h+\frac{1}{2}\left\{h,h\right\}\right) - \\
\int_{M}\d^{4|8}x \left[\int_{X}Y_{I}\left(\dbar_{\sigma}Z^{I}-I^{IJ}\partial_{J}h\right)+\frac{\varepsilon^{2}}{\Lambda \kappa^{2}}\left(\int_{X} \tilde{h}\wedge\tau\right)^2 \right].
\end{multline}
We do not currently have a direct proof (analogous to theorem \ref{MThm} for conformal gravity) that this action corresponds to Einstein gravity.  In this paper, we justify its validity by the calculations that follow and by the embedding of Einstein gravity in conformal gravity.  In particular, \eqref{EinTA4} is obtained from the conformal gravity twistor action by simply restricting to Einstein degrees of freedom and then applying Maldacena's argument.  Furthermore, when $\varepsilon= 0$ it reduces to the correct twistor action for self-dual Einstein gravity \cite{Mason:2007ct}, and in the next section, we show that the non-self-dual interactions produce the correct MHV amplitude.

It is worth noting that we can easily define similar actions for $\cN=0$ and $\cN=8$ Einstein gravity.  In the first case, the two gravitons of general relativity are given (off-shell) in twistor space by $\tilde{h}\in\Omega^{0,1}_{\CPT}(-6)$, $h\in\Omega^{0,1}_{\CPT}(2)$.  The resulting twistor action is then
\begin{multline}\label{EinTA0}
S^{\mathrm{Ein}}_{\cN=0}[\tilde{h},h]=\int_{\CPT}\D^{3}Z\wedge\tilde{h}\wedge\left(\dbar h+\frac{1}{2}\left\{h,h\right\}\right) - \\
\int_{M}\d^{4}x \left[\int_{X}Y_{\alpha}\left(\dbar_{\sigma}Z^{\alpha}-I^{\alpha\beta}\partial_{\beta}h\right)+\frac{\varepsilon^{2}}{\Lambda \kappa^{2}}\int_{X\times X} \la\lambda_{1}\lambda_{2}\ra^{4}\;\tilde{h}_{1}\wedge\tau_{1}\wedge\tilde{h}_{2}\wedge\tau_{2}\right].
\end{multline}
For $\cN=8$ supersymmetry, twistor space is topologically $\CP^{3|8}$ and the single graviton multiplet is encoded by $h\in\Omega^{0,1}_{\CPT}(2)$, which incorporates the negative helicity graviton in the term $\chi^{8}\tilde{h}$.  This leads to an action:
\begin{multline}\label{EinTA8}
S^{\mathrm{Ein}}_{\cN=8}[h]=\int_{\CPT}\D^{3|8}Z\wedge h\wedge\left(\dbar h+\frac{1}{3}\left\{h,h\right\}\right) - \\
\int_{M}\d^{4|16}x \left[\int_{X}Y_{I}\left(\dbar_{\sigma}Z^{I}-I^{IJ}\partial_{J}h\right)+\frac{\varepsilon^{2}}{\Lambda \kappa^{2}}\int_{X\times X} \frac{h_{1}\wedge\tau_{1}\wedge h_{2}\wedge\tau_{2}}{\la\lambda_{1}\lambda_{2}\ra^{4}}\right].
\end{multline}


\section{Perturbation theory and the MHV Amplitude}
\label{MACC}

In this section, we consider the perturbation theory associated to our twistor actions, and use it to derive a formula for the MHV tree amplitude of Einstein gravity in the presence of a cosmological constant, $\Lambda$ in twistor space. This provides a check on our claim that the twistor action \eqref{EinTA4} describes Einstein gravity.   

We first make some general remarks about the perturbation theory for the twistor actions in an axial gauge (otherwise known as the CSW gauge in the case of Yang-Mills, \cite{Cachazo:2004kj}).  Our discussion here will take place in twistor space, and our development of the Feynman diagrams will be restricted to those that are required for the MHV amplitude.   We discuss the extension to more general  amplitudes in Section \ref{DC}.

For the conformal gravity twistor action \eqref{CGTA}, the axial gauge is a choice of coordinates and gauge for $g$ so that one of the anti-holomorphic form components of $f$ and $g$ vanish, say in the direction of some fixed choice of reference twistor $Z_*$:
\be{CSW-gauge}
\overline{Z_*\cdot\frac{\partial}{\partial Z}} \lrcorner f=0=\overline{Z_*\cdot\frac{\partial}{\partial Z}} \lrcorner g  \, ,
\ee 
with identical restrictions on $h$ and $\tilde h$ in the Einstein case.
This has the effect of eliminating the cubic term in the self-dual
part of the twistor action, so all remaining vertices are in the non-self-dual interaction terms.  In the case of conformal gravity, these vertices can be read off from \eqref{minTA3}:
\be{Verts0}
V_{f}=\int_{M}\d^{4|8}x\int_{X}Y_{I} f^{I}, \qquad V_{g}=\int_{M}\d^{4|8}x\int_{X\times X}g_{1}\wedge g_{2}.
\ee

The axial gauge leaves us with two kinetic terms: $g_{I}\dbar f^{I}$ (or $\tilde h\dbar h$) from the self-dual portion of the action, and $Y_{I}\dbar_\sigma Z^{I}$ from the second part.  The first lives on twistor space and the second on the Riemann sphere $X\cong\CP^1$, so they lead to propagators $\Delta(Z,Z')$ on twistor space connecting an $f$ and a $g$ (or a $h$ and a $\tilde h$), and $\Delta (\sigma,\sigma')$  on $X$ connecting a $Y$ and a $Z$.

The MHV degree $k$ of a tree amplitude is the count of the number of external $g$s (or $\tilde h$s) in an amplitude minus 2.  Since each $V_g$ inserts two $g$s, and each propagator $\Delta(Z,Z')$ takes the place of one $g$, we have
\be{MHV-deg}
k= |V_g|+l-1
\ee
where $|V_{g}|$ is the number of $V_g$ insertions and $l$ is the number of loops in the diagram obtained by deleting all the propagators on the Riemann sphere (this follows because $l=1+|\Delta(Z,Z')|-|V_g|$).  For the rest of this section we will be working at tree-level, and concerned with computing the MHV amplitude.  This means that we need only consider a single $V_g$ vertex with no $\Delta(Z,Z')$ propagators in play.

While we will focus on Einstein states in the following, much of our calculation is easily applicable to conformal gravity since polarization states for this theory can be expressed in terms of Einstein states with different conformal factors.  Given the permutation symmetry of the positive helicity and negative helicity fields amongst themselves, we can generate all conformal gravity amplitudes by considering Einstein states with one choice of infinity twistor for the positive helicity states (upstairs indices), and a different one for the negative helicity states (downstairs indices).  Restricting to the Einstein subsector is then accomplished by requiring these infinity twistors be compatible as in \eqref{inftyCC}.


\subsection{A Feynman diagram calculus for the MHV amplitude}

In this subsection we obtain the Feynman diagrams that contribute to the MHV amplitude and in the next sum them using the matrix-tree theorem
to give a compact formula in terms of reduced determinants analogous to that of Hodges \cite{Hodges:2012ym}.  This explains the use of the matrix-tree theorem for this amplitude as first described in \cite{Feng:2012sy, Adamo:2012xe} and elaborated below.

At MHV, and in the axial gauge, the only part of the twistor action which is relevant for the computation is 
\be{Vertgen2}
\int_{M}\d^{4|8}x \left[\int_{X}\left(Y_{I}\dbar_{\sigma}Z^I+ [Y,\p
    h]\right)+\int_{X\times X}\tilde{h}_{1}\;\tau_{1}\wedge\tilde{h}_{2}\;\tau_{2}\right],
\ee
where we have included the Lagrange multiplier field $Y_{I}$ as in \eqref{minTA3}.  Note that since there are no $\Delta(Z,Z')$ propagators, the perturbation theory associated to \eqref{Vertgen2} is the same for both twistor actions and is also equivalent to the perturbative expansion of the MHV generating functional \eqref{CGDS*}.

We consider only Feynman tree diagrams, and read off from the action the vertices: 
\be{Vertices}
V_h= \int_{M}\d^{4|8}x \int_X [Y,\p h] \, , \qquad V_{\tilde
  h}=\int \rd^{4|8}x \int_{X\times X}\tilde{h}_{1}\;\tau_{1}\wedge\tilde{h}_{2}\;\tau_{2}.
\ee
We only use the propagator for the $Y$-$Z$ kinetic term, which is fixed to coincide with \eqref{Picard2}:
\be{prop}
\left\la Y_{I}(\sigma_{i})\;Z^{J}(\sigma_j)\right\ra=\Delta(\sigma_i,\sigma_j)=\frac{(\xi \sigma_j)^2 \D\sigma_i}{(\xi\sigma_i)^2 (\sigma_i
  \sigma_j)}=\frac{(\xi j)^2 \D\sigma_i}{(\xi i)^2 (i \, j)},
\ee
suppressing the delta function in $x$ (which just restricts the calculation to the
Riemann sphere $X$ leaving one overall integral over space-time).

It follows from the remarks above that the Feynman diagrams contributing to the $n$-point MHV amplitude are those in which there is one $V_{\tilde h}$ vertex, which contains two external wavefunctions (the negative helicity $\tilde h$s), and $n-2$ $V_h$ vertices.  On-shell $Y$ vanishes, being holomorphic on $\CP^1$ of negative weight, so the $n-2$ $Y$s (one in each of the $V_h$ vertices) must each be contracted via a propagator $\Delta(\sigma, \sigma')$.  These in turn connect with $Z$s, which occur in the $h$s, $\tilde{h}$s, and $\tau$s of the other vertices.

It is convenient to expand the vertex $V_{\tilde h}$ into four separate vertices in the diagram, each corresponding to the four sites of $Z$-dependence that a $Y$-$Z$ propagator can attach itself to.  Thus, the Feynman diagram calculus on $\CP^1$ for the $n^{\mathrm{th}}$-order perturbative evaluation of the generating functional \eqref{Vertgen2} is defined as follows:  
\begin{itemize}
\item Draw a black vertex for each of $\tilde{h}_{1}$, $\tilde{h}_{2}$.

\item Draw a grey vertex for each contact structure $\tau_{1}$, $\tau_{2}$. 

\item Draw a white vertex for each of the $n-2$ vertices $V_{h_{i}}$, $i=3, \ldots , n-2$.

\item Draw an oriented edge out from each white vertex to some other vertex such that the resulting diagram is a forest of trees rooted at a black or grey vertex.
\end{itemize}
\begin{figure}[h]
\centering
\includegraphics[width=2.1 in, height=0.35 in]{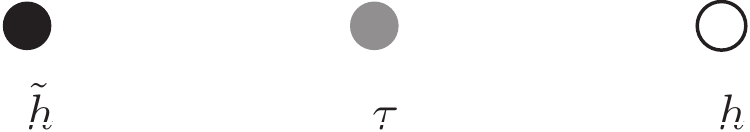}\caption{\small{\textit{Building blocks for Feynman diagrams}}}\label{FeynRules}
\end{figure}

Each diagram corresponds to an integrand to be integrated over the $n$-fold product of the $\CP^1$ factor in \eqref{Vertgen2} and then the
final expression must be integrated against the measure $\d^{4|8}x$.  The vertices are each associated to a point $\sigma_i$ on the $i^{\mathrm{th}}$ $\CP^1$
factor.  For $i=1,2$ we have
$$
h_i:=\tilde
h_i(Z(\sigma_i))\, , \qquad \tau_i:= I_{IJ}Z^I(\sigma_i)\p
Z^J(\sigma_i)
$$ 
at the black and grey vertices respectively.  Writing
\begin{equation*}
Z_i=Z(\sigma_i)\, , \quad \p_{iI}=\frac{\p}{\p Z^I(\sigma_i)}\, , \quad Y_{iI}=Y_I(\sigma_i)\, ,\quad h_i= h_i(Z(\sigma_i)),
\end{equation*}
(and often suppressing the $I,J$ indices) we obtain the $n-2$ white
vertices of the form $\int [Y_j,\partial_j h_j]$.  The kinetic term
$Y_{I}\dbar_{\sigma}Z^{I}$ defines the propagator \eqref{prop}, and
the removal of a $Z$ from a vertex to replace with the end of the
propagator corresponds to differentiation with respect to $Z$.  Thus an edge from a white node $j>2$ to a black or white node $i$ corresponds to the differential operator
\be{edge}
\frac{(\xi \sigma_i)^2\; \D\sigma_j}{(\xi\sigma_j)^2 (\sigma_j
  \sigma_i)}
[\p_j h_j,\p_i]
\ee
acting on the wave function at the $i^{\mathrm{th}}$ node of the
diagram.  We give the action on $\tau$ at a grey vertex below.

Since there is a single $Y_{I}$ in each white vertex, there are $n-2$
total edges in each diagram.  The wavefunctions $\tilde{h}$, $h$ depend non-polynomially on $Z$, so the white and black vertices can have have an arbitrary number of incoming edges.  Since $\tau=\la Z(\sigma),\partial Z(\sigma)\ra$ is of order two in $Z$, the grey vertices can absorb at most two edges.

To summarize, we represent the perturbative expansion of the MHV generating functional \eqref{Vertgen2} by using a $\CP^{1}$-Feynman diagram calculus.  Since we work classically, each diagram corresponds to a forest of trees on $n+2$ (2 $\tau$s + 2 $\tilde{h}$s + $n-2$ $h$s) vertices, rooted at a black or grey vertex.



\subsubsection*{\textit{Propagators ending on} $\tau$}

Each diagram has two grey vertices corresponding to the contact structures $\tau_{i} = \la Z_{i},\partial Z_{i}\ra$,
$i=1,2$ in the $V_{\tilde h}$ vertex.  These are quadratic in $Z$ and so can have at most two incoming arrows; higher numbers of incoming arrows will vanish.  If the upstairs infinity twistor is the inverse of the downstairs one (as in the Einstein case), other contributions vanish as follows.

\begin{figure}[h]
\centering
\includegraphics[width=3.20 in, height=0.3 in]{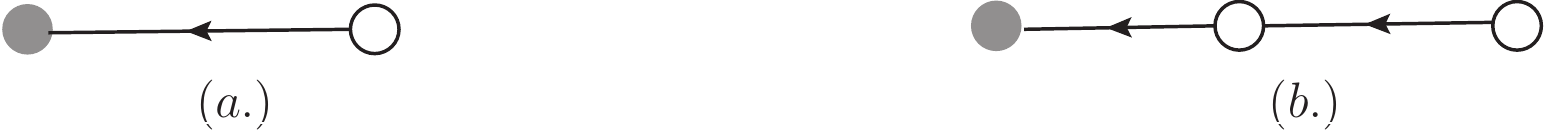}\caption{\small{\textit{An isolated \emph{(a.)} and un-isolated \emph{(b.)} component}.}}\label{taus}
\end{figure}

\begin{lemma}\label{taucont}
If a Feynman diagram has a disconnected piece with just one white vertex connected to a grey vertex, we refer to it as \emph{isolated} if the corresponding white vertex has no incoming arrows, as in \emph{(a.)} of Figure \ref{taus}.  An isolated disconnected piece in a diagram leads to a factor which vanishes after integration by parts.  In particular, an isolated propagator connecting a white vertex $i$ to a grey vertex $1$ produces a factor 
$$
 2\Lambda \D\sigma_{1}\; \sigma_{1A}\int_{\CP^1}\rd_{i} \left(\frac{\sigma^{A}_{i}(\xi 1)h_{i}}{(1i)^{2}(\xi i)}\right),
$$
where $\rd_i$ is the exterior derivative in the $\sigma_i$ variable and integration by parts makes the contribution vanish.
\end{lemma}

The proof follows by observing that when the $i^{\mathrm{th}}$ white node is connected to $Z_1$ we must replace
\be{Picit1}
Z^{I}(x,\sigma_1)\rightarrow\int_{X_i}\frac{\D\sigma_i}{(1
  i)}\frac{(\xi 1)^2}{(\xi i)^2}I^{IJ}\partial_J h_i\, .
\ee
The lemma results from a direct but slightly tedious computation which is relegated to Appendix \ref{taucontapp}.  

Thus we can neglect any \emph{isolated} arrows to the contact structures in our diagrams.  However,  if $\tau_{1}$ is connected to vertex $i$ which is in turn connected to vertex $j$, then additional $\sigma_{i}$-dependence is introduced by the propagator from the $j^{\mathrm{th}}$ vertex and we no longer obtain the total derivative in lemma \ref{taucont}.  Similarly, we eliminated the second $Z_1$ in $\tau$ using the linear dependence of $Z$ on $\sigma$ but a second propagator could have been inserted there, so for higher order contributions we will need further calculation as follows.

When a propagator connects to a grey vertex, there is always a contraction between the upstairs and downstairs infinity twistors so we obtain a factor of $\Lambda$.  After a bit of algebra, we find a single propagator plugged into the contact structure (say, $\tau_1$) is given by:
\be{tauprop1}
\psi^{1}_{i}=
\Lambda \frac{\D\sigma_i (\xi 1)^4}{(1i)^{2}(\xi i)^2} \rd_1\left( \frac{ (i1)}{(\xi 1)^2}Z^I_1\right)\partial_{iI} h_i 
=\Lambda \frac{\D\sigma_1 \D\sigma_i (\xi 1)}{(1i)^{2}(\xi i)^2}\left[(\xi i)\;Z^{I}_{1}+(1i)\;Z^{I}(\xi)\right]\partial_{iI} h_i \, ,
\ee
where the first formula must be used when further propagator insertions are required, as the second uses the linearity of $Z$ as a function of $\sigma$.
Similarly, for two propagators plugged into the contact structure we obtain:
\be{tauprop2}
\omega^{1}_{ij}=-\Lambda\frac{\D\sigma_1 \D\sigma_i \D\sigma_j (1\xi)^{4}(ij)}{(1i)^{2}(1j)^{2}(\xi i)^{2}(\xi j)^{2}}\left[\partial_i,\partial_j \right] h_i h_j \, .
\ee
Note that there are many equivalent formulae for these following from the Schouten identity but these are what we will use in the following calculations.
\begin{figure}[t]
\centering
\includegraphics[width=4.20 in, height=2 in]{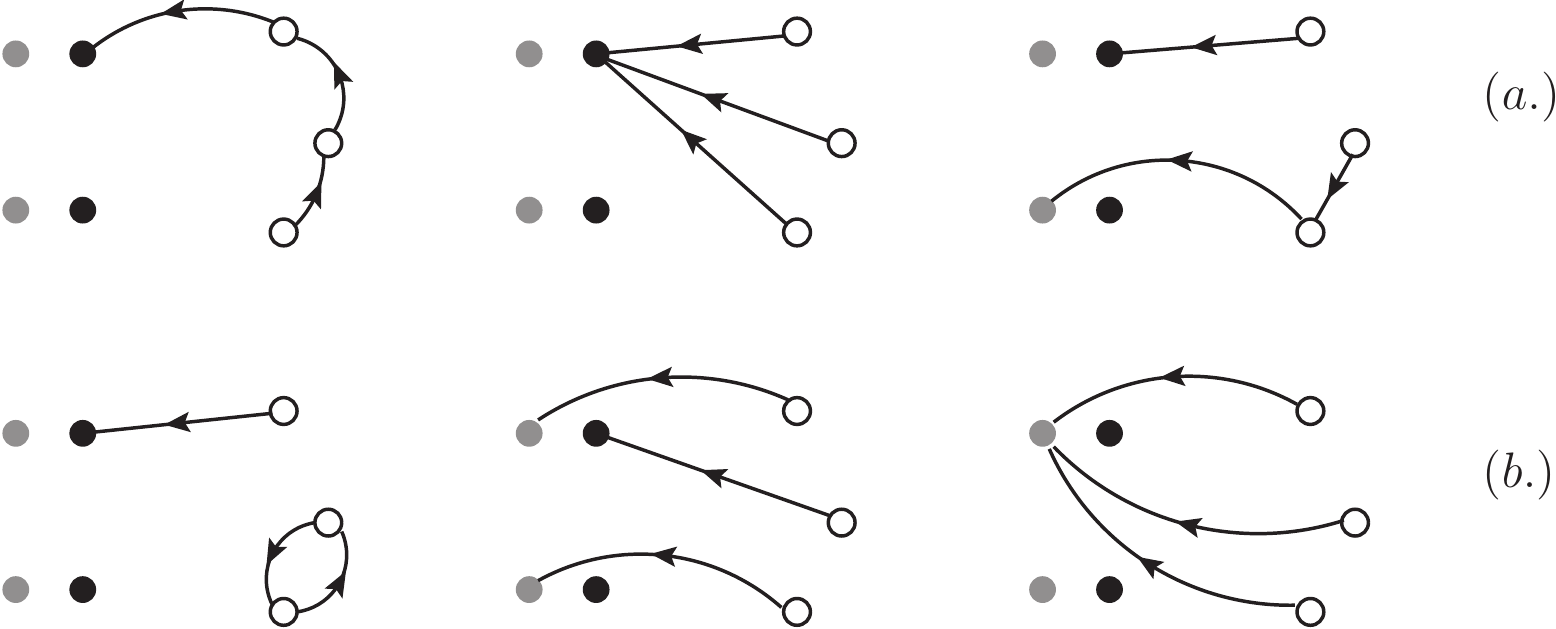}\caption{\small{\textit{Some diagrams for the 5-point amplitude which have a non-vanishing \emph{(a.)}, or excluded/vanishing \emph{(b.)} contribution.}}}\label{Diags}
\end{figure}

Clearly, at any order in $n$ there are many diagrams which can be drawn on the $n+2$ vertices which are either excluded or vanish.  In Figure \ref{Diags}, we illustrate several examples for the case of the 5-point amplitude.  All the diagrams in (\emph{a.}) give a non-vanishing contribution, while all those in (\emph{b.}) are either excluded or vanish.  In the latter case, the first diagram of (\emph{b.}) is excluded because of the loop; the second vanishes because there are isolated propagators to the contact structure so lemma \ref{taucont} applies; and the third vanishes because there are more than two propagators ending in a contact structure.


\subsection{Matrix tree formulae for the MHV amplitude}

For a $n$-point MHV diagram, we must sum all of the associated $\CP^1$ Feynman diagrams as described above.  Each of these diagrams will have the structure of a set of tree graphs (i.e., a forest) on $n+2$ vertices, where each tree is rooted at one of the black or grey vertices and the edges/arrows correspond to propagators.  The most efficient way to perform this sum is by using a powerful result from algebraic combinatorics known as the \emph{matrix-tree theorem} (textbook treatments of can be found in \cite{Stanley:1999, vanLint:2001, Stanley:2012}).

Let us review this theorem in the context of an arbitrary graph.  Suppose $G$ is some oriented graph with a set of $n$ vertices $\mathcal{V}=\{i\}_{i=1,\ldots,n}$ and edges $\mathcal{E}$, where an edge connecting vertex $i$ to vertex $j$ is denoted by $(i,j)\in\mathcal{E}$.  We can promote $G$ to a weighted graph by endowing each edge $(i,j)$ with a weight $w_{ij}\in\C$.  This data is naturally encoded into the \emph{weighted Laplacian matrix} of $G$, which is a $n\times n$ matrix with entries:
\begin{equation*}
\mathcal{L}_{ij}(G)=\left\{
\begin{array}{c}
-w_{ij} \;\mathrm{if}\;i\neq j\;\mathrm{and}\;(i,j)\in\mathcal{E}\\
\sum_{(i,k)\in\mathcal{E}}w_{ik}\;\mathrm{if}\;i=j \\
0 \;\;\mathrm{otherwise}
\end{array}\right. .
\end{equation*}  

The weighted Laplacian matrix is the basic ingredient in the matrix-tree theorem, which tells us how to count tree subgraphs of $G$ including weights.  A \emph{tree of} $G$ \emph{rooted at} $i\in\mathcal{V}$ is a sub-graph $T^{(i)}=(\mathcal{V},E^{(i)})$ of $G$ such that: (1.) $T^{(i)}$ has no oriented cycles; (2.) vertex $i$ has no outgoing edges; and (3.) every other vertex $j\neq i$ has one outgoing edge.  We have denoted the edges of the tree $T^{(i)}$ by $E^{(i)}\subset\mathcal{E}$.  A collection of $r$ rooted trees of $G$ is called a \emph{rooted forest}; we will denote the set of such forests rooted at vertices $\{i_{1},\ldots,i_{r}\}$ by $\mathcal{F}^{(i_1,\ldots, i_r)}(G)$.

The matrix-tree theorem for rooted forests on the directed graph $G$ is then given by:
\begin{thm}[Weighted Matrix-Tree Theorem for Forests]\label{MTT}
Let $\mathcal{F}^{(i_1,\ldots i_r)}(G)$ be the set of forests of $G$ rooted at $\{i_1,\ldots, i_r\}\subset\mathcal{V}$ and $\mathcal{L}(G)$ be the weighted Laplacian matrix of $G$.  For each $F\in\mathcal{F}^{(i_1,\ldots i_r)}(G)$, denote by $E_F\subset\mathcal{E}$ the set of edges in the forest.  Then
\be{MTT*}
\left|\mathcal{L}(G)^{i_{1}\cdots i_{r}}_{i_{1}\cdots i_{r}}\right|= \sum_{F\in\mathcal{F}^{(i_1,\ldots i_r)}(G)}\left(\prod_{(i,j)\in E_{F}}w_{ij}\right),
\ee
where $ \left|\mathcal{L}(G)^{a\cdots b}_{c\cdots d}\right|$ denotes the determinant of $\cL(G)$ with the rows $\{a,\ldots, b\}$ and columns $\{c,\ldots, d\}$ removed.
\end{thm}
A proof of this particular version of the matrix-tree theorem can be found in \cite{Feng:2012sy}. 

\medskip

For the situation we are interested in, $G$ is the graph on $n+2$ vertices (2 grey, 2 black, and $n-2$ white) with \emph{all} possible propagator edges drawn in.  The weights for edges between white and black or white and white vertices correspond to the propagators \eqref{edge}:
\begin{equation*}
w_{ij}\leftrightarrow \frac{(\xi j)^{2}\;\D\sigma_{i}}{(\xi i)^{2}(ij)}\left[\partial_{i}h_i, \partial_{j}\right],
\end{equation*}
and edges to a grey vertex correspond to the propagators \eqref{tauprop1} or \eqref{tauprop2}.  This clearly constitutes all the data required to build the weighted Laplacian matrix for $G$ and then apply theorem \ref{MTT}.

To proceed, denote the set of all Feynman diagrams contributing to the $n$-point amplitude as $\mathcal{F}^{n}$.  This set has a natural disjoint-union splitting based upon the number of arrows which are incoming at each of the two contact structures (grey vertices) $\tau_{1}$, $\tau_{2}$.  Explicitly, we have
\begin{equation*}
\mathcal{F}^{n}=\bigsqcup_{k=0}^{4}\mathcal{F}^{n}_{k},
\end{equation*}
where each diagram $\Gamma\in\mathcal{F}^{n}_{k}$ is a forest on $n+2$ vertices which has $k$ arrows into the contact structures (for $k>4$ all the diagrams have a vanishing contribution since $\tau$ is quadratic in $Z$).

The simplest case involves no propagators to the contact structures, where we can write $\tau_{1,2}=X^{2}\;\D\sigma_{1,2}$ for $X^{2}\equiv I_{IJ}X^{I}_{A}X^{JA}$.  The contribution to the $n$-point vertex can be written as
\begin{equation*}
\sum_{\Gamma\in\mathcal{F}^{n}_{0}}\int \d^{4|8}x\;(X^{2})^{2}\;F_{\Gamma}\;\prod_{i=1}^{n}h_{i}\;\D\sigma_{i},
\end{equation*}
where $F_{\Gamma}$ encodes the contribution from diagram $\Gamma$ built out of the propagators, which are all of the form \eqref{edge}.\footnote{Here, we think of the integral as being over $\M\times (\CP^1)^n$ but this can also be thought of as an integral over $\CM_{n,1}$, the moduli space of $n$-pointed holomorphic maps $Z^{I}:\CP^{1}\rightarrow\PT$ of degree one \cite{Adamo:2012cd}.}

Since there are no propagators to the grey vertices, each term in this sum corresponds to a forest of trees rooted at the two black vertices corresponding to $\tilde{h}_{1}$ and $\tilde{h}_{2}$.  We can perform this sum using theorem \ref{MTT} after constructing the weighted Laplacian matrix for the remaining propagators, as was first shown in \cite{Feng:2012sy, Adamo:2012xe}.  Up to an irrelevant conjugation, this weighted Laplacian matrix takes the form of the `Hodges matrix' $\HH$ whose entries are given by:
\be{HodgesMatrix}
\HH_{ij}=\left\{
\begin{array}{c}
\frac{1}{(ij)}\left[\partial_{i}, \partial_{j}\right] \:\mbox{if}\: i\neq j \\
-\sum_{j\neq i}\HH_{ij}\frac{(\xi j)^{2}}{(\xi i)^2} \:\mbox{if}\: i=j
\end{array}\right. ,
\ee
in accordance with \eqref{edge}.  Note that this means the entries in our Laplacian matrix take the form of differential operators which will (after applying the matrix-tree theorem below) act on the wavefunctions $\{h_{i}\}$.  With momentum eigenstates and a generic infinity twistor these operators become rather complicated, involving derivatives of delta-functions.  Our manipulations would be considerably simpler if we could treat these terms algebraically.  

This can be accomplished by working with \emph{dual twistor} wavefunctions:
\be{dtwf}
h(Z(\sigma_{i}))=\int_{\C}\frac{\d t_{i}}{t_{i}^{1+w_{i}}}\exp\left(i t_{i}W_{i}\cdot Z(\sigma_{i})\right), \qquad w_{i}=\left\{
\begin{array}{cc}
-2 & \mbox{if}\;i=1, 2 \\
2 & \mbox{otherwise}
\end{array} \right. .
\ee
Here $W_{i\;I}=(\tilde{\mu}^{A}, \tilde{\lambda}_{i A'})$ are coordinates on $n$ copies of dual twistor space, $\PT^{\vee}$.  These wavefunctions have been used before in other contexts \cite{Mason:2009sa, Cachazo:2012pz}, and can be paired with momentum eigenstates in an appropriate manner to obtain functionals of momenta at the end of any calculation.  Furthermore, the scaling parameters $t_{i}$ can be absorbed into the worldsheet coordinates by defining a new set of non-homogeneous coordinates: $\sigma_{i}t_{i}\rightarrow\sigma_{i}$, $\d t_{i}\D\sigma_{i}\rightarrow\d^{2}\sigma_{i}$.

With \eqref{dtwf}, all the propagators of the Feynman diagram calculus become purely algebraic.  In particular, we now have:
\begin{equation*}
\HH_{ij}=-\frac{[W_{i},W_{j}]}{(ij)},
\end{equation*}
so the weighted Laplacian matrix \eqref{HodgesMatrix} becomes an algebraic object.  The propagators involving the contact structure also become algebraic in this picture:
\begin{equation*}
\psi^{1}_{i}=\Lambda\;i\frac{(\xi 1)\;W_{i\;I}}{(1i)^{2}(\xi i)^2}\left[(\xi i)\;Z^{I}(\sigma_{1})+(1i)\;Z^{I}(\xi)\right], \qquad \omega^{1}_{ij}=\Lambda \frac{[W_{i},W_{j}]\;(1\xi)^{4}(ij)}{(1i)^{2}(1j)^{2}(\xi i)^{2}(\xi j)^{2}}.
\end{equation*}
Hence, we always have the option of moving from generic twistor states, where the propagators and Laplacian matrix take the form of differential operators, to dual twistor states where they become algebraic quantities.

Returning to the sum of Feynman diagrams in $\mathcal{F}^{n}_{0}$, we apply theorem \ref{MTT} to obtain the contribution
\be{cont0}
\int \d^{4|8}x\;(X^{2})^{2}\;\left| \HH^{12}_{12}\right|\;\prod_{i=1}^{n}h_{i}\;\D\sigma_{i}.
\ee
The notation $| \HH^{12}_{12}|$ indicates the determinant of $\HH$ with the row and columns corresponding to $\tilde{h}_{1}$ and $\tilde{h}_2$ removed.

We can now apply the matrix-tree theorem in a similar fashion to the other subsets of Feynman graphs $\mathcal{F}^{n}_{k>0}$.  For instance, consider graphs in $\mathcal{F}^{n}_{1}$.  The single deformation of the contact structure may come from any white vertex $i=3,\ldots,n$, and results in a propagator $\psi^{1}_{i}$ or $\psi^{2}_{i}$ from \eqref{tauprop1}.  All the remaining arrows in the graph will correspond to propagators captured by the weighted Laplacian matrix \eqref{HodgesMatrix}, so once we factor out the propagator to $\tau$ we are in the business of counting forests of trees rooted at vertices $1$, $2$, or $i$.  Via theorem \ref{MTT}, we then have:
\be{cont1}
\sum_{\Gamma\in\mathcal{F}^{n}_{1}}\int \d^{4|8}x\;X^{2}\;F_{\Gamma}\;\prod_{i=1}^{n}h_{i}\;\D\sigma_{i} =\int \d^{4|8}x\;X^{2}\sum_{i=3}^{n}\psi^{1}_{i}\;\left| \HH^{12i}_{12i}\right|\; \prod_{j=1}^{n}h_{j}\;\D\sigma_{j}+(1\leftrightarrow 2).
\ee
Note that there is only a single power of $X^{2}$ appearing as an overall factor in this expression; the other has been eaten by the propagator $\psi^{1}_{i}$.  Also recall that although $\psi^{1}_{i}$ and the entries of $\HH$ are generally differential operators, we can think of them as algebraic quantities by working with the dual twistor wavefunctions \eqref{dtwf}.

A similar pattern follows for the remaining subsets in $\mathcal{F}^{n}$.  Adding all of them together and including the required factor of $\Lambda^{-1}$ from the embedding of Einstein gravity into conformal gravity gives us the following formula for the MHV amplitude:
\begin{multline}\label{MHVamp}
\cM_{n,0}=\frac{1}{\Lambda}\int \d^{4|8}x\;\left[ (X^2)^2\left|\HH^{12}_{12}\right|+ X^{2} \sum_{i}\psi^{1}_{i}\left|\HH^{12i}_{12i}\right| +X^{2}\sum_{i,j}\omega^{1}_{ij}\left|\HH^{12ij}_{12ij}\right| \right. \\
\left. +\sum_{i,j}\psi^{1}_{i}\psi^{2}_{j}\left|\HH^{12ij}_{12ij}\right| +\sum_{i,j,k}\psi^{1}_{i}\omega^{2}_{jk}\left|\HH^{12ijk}_{12ijk}\right| +\sum_{i,j,k,l}\omega^{1}_{ij}\omega^{2}_{kl}\left|\HH^{12ijkl}_{12ijkl}\right|\right]\prod_{m=1}^{n}h_{m}\;\D\sigma_{m}\:+(1\leftrightarrow 2).
\end{multline}
In this expression, the sums are understood to run over all indices which are not excluded from the determinant, and also to symmetrize on those indices.  For instance, in the first term of the second line $\sum_{i,j}$ runs over all $i,j=3,\ldots n$ with $i\neq j$.  

This formula is a perfectly valid representation of the MHV amplitude with cosmological constant; it can be simplified substantially if we investigate its properties a bit further, however.  To do this, we adopt the dual twistor wavefunctions \eqref{dtwf} so that all propagators become algebraic.  Working with the re-scaled coordinates $\sigma_{i}t_{i}\rightarrow\sigma_{i}$, the product of wavefunctions and measures can be expressed compactly as    
\begin{equation*}
\prod_{i=1}^{n}h_{i}\;\D\sigma_{i}=e^{i\mathcal{P}\cdot X}\;\d^{2}\sigma, \qquad \cP_{I}^{A}=\sum_{i=1}^{n}W_{i\;I}\sigma_{i}^{A}, \qquad \d^{2}\sigma\equiv \prod_{i=1}^{n}\d^{2}\sigma_{i}.
\end{equation*}

Now note that the second term in the first line of \eqref{MHVamp} can be written as
\begin{multline*}
\int\d^{4|8}x\;X^{2}\sum_{i} \psi^{1}_{i}\left|\HH^{12i}_{12i}\right|\;e^{i\cP\cdot X}\d^{2}\sigma \\
=\Lambda \int \d^{4|8}x\;X^{2}\sum_{i}\left|\HH^{12i}_{12i}\right|\left(\frac{(\xi 1)(\xi i)\sigma_{1}^{A}+(\xi 1)(1i)\xi^{A}}{(1i)^{2}(\xi i)^{2}}\right)\frac{\partial e^{i\cP\cdot X}}{\partial\sigma_{i}^{A}}\d^{2}\sigma ,
\end{multline*}
where we have used the algebraic expression for $\psi^{1}_{i}$.  This is expected as a result of lemma \ref{taucont}, which tells us that a propagator to one of the contact structures takes the form of a derivative with respect to $\sigma_{i}$.  Hence, we can integrate by parts with respect to $\d^{2}\sigma_{i}$ to find:
\begin{multline*}
\int\d^{4|8}x\;X^{2}\sum_{i} \psi^{1}_{i}\left|\HH^{12i}_{12i}\right|\;e^{i\cP\cdot X}\d^{2}\sigma  \\
=-\Lambda\int\d^{4|8}x\;X^{2}e^{i\cP\cdot X}\sum_{i}\frac{\partial}{\partial\sigma_{i}^{A}}\left(\left|\HH^{12i}_{12i}\right|\frac{(\xi 1)(\xi i)\sigma_{1}^{A}+(\xi 1)(1i)\xi^{A}}{(1i)^2(\xi i)^2}\right) \d^{2}\sigma \\
=-\Lambda\int\d^{4|8}x\;X^{2}e^{i\cP\cdot X}\sum_{i,j}\left|\HH^{12ij}_{12ij}\right|\frac{[W_{i},W_{j}](1\xi)^{4}(ij)}{(1i)^{2}(1j)^{2}(\xi i)^{2}(\xi j)^{2}}\d^{2}\sigma \\
=-\int\d^{4|8}x\;X^{2}\sum_{i,j}\omega^{1}_{ij}\left|\HH^{12ij}_{12ij}\right| e^{i\cP\cdot X}\d^{2}\sigma .
\end{multline*}
with the third line following after symmetrizing over $(i\leftrightarrow j)$ and several applications of the Schouten identity.

Thus, we see that following an integration by parts the second term in \eqref{MHVamp} cancels the third term.  A similar calculation demonstrates that the fourth and fifth terms also cancel with each other.  We are therefore able to reduce our formula for the amplitude to one with only two terms:
\be{MHVamp2}
\cM_{n,0}=\frac{1}{\Lambda}\int\d^{4|8}x\;\left[ (X^2)^2\left|\HH^{12}_{12}\right| +\sum_{i,j,k,l}\omega^{1}_{ij}\omega^{2}_{kl}\left|\HH^{12ijkl}_{12ijkl}\right|\right]\prod_{m=1}^{n}h_{m}\;\D\sigma_{m}\:+(1\leftrightarrow 2),
\ee
where we have restored arbitrary twistor wavefunctions and homogeneous coordinates.  Clearly this formulation is an improvement over \eqref{MHVamp} in terms of simplicity.

A non-trivial test which any formula for $\cM_{n,0}$ must pass is that it must be independent of the reference spinor $\xi\in\CP^{1}$.  This entered the definition of the propagator $\Delta(\sigma,\sigma')$ due to the ambiguity in defining $\dbar^{-1}_{\sigma}$ on forms of positive degree.  Hence, the choice of $\xi$ is equivalent to a choice of gauge for the propagator on $\CP^1$; by \eqref{Picard3} a variation in $\xi$ should correspond to a diffeomorphism on the projective spinor bundle $\PS$.  In other words, observables such as $\cM_{n,0}$ should be independent of the reference spinor.

An obvious way of demonstrating this is to consider the infinitesimal variation generated by the derivative $\d_{\xi}=\d\xi^{A}\frac{\partial}{\partial\xi^{A}}$.  The calculation of $\d_{\xi}\cM_{n,0}$ is a lengthy but relatively straightforward procedure which is carried out in Appendix \ref{XI}; the final result is that
\be{xivar2}
\d_{\xi}\cM_{n,0}=\int\frac{\d^{8|8}X}{\mathrm{vol}\;\GL(2,\C)}\frac{\partial}{\partial X^{IA}} V^{IA}=0,
\ee
where $V^{IA}$ are the components of a smooth vector field (roughly speaking, on $\CM_{n,1}$).  The fact that $\d_{\xi}\cM_{n,0}$ vanishes as a total divergence indicates that a variation in $\xi$ corresponds to a diffeomorphism on the spinor bundle $\PS$, and proves that \eqref{MHVamp}, \eqref{MHVamp2} is a well-defined formula for the amplitude.


\subsection{The flat-space limit and the Hodges formula}

A final test which our expression for $\cM_{n,0}$ must pass is the flat-space limit, where it should reproduce Hodges' formula for the MHV amplitude \cite{Hodges:2012ym}.  In the language of $\cN=4$ supergravity, Hodges' formula is:\footnote{Note that there are many equivalent representations of this formula, we have simply presented the one which connects most directly to our conformal gravity arguments.}
\be{HForm}
\cM^{\mathrm{Hodges}}_{n,0}(\Lambda=0)=\int \d^{4|8}x\; \frac{(12)^{2}}{(1i)^{2}(2i)^{2}}\left|\HH^{12i}_{12i}\right|\prod_{j=1}^{n}h_{j}\;\D\sigma_{j}.
\ee

Initially, it appears that the structure of $\cM_{n,0}$ is a long way off from Hodges' formula.  If we use dual twistor wavefunctions \eqref{dtwf}, then \eqref{MHVamp2} takes the form
\be{3red1}
\cM_{n,0}=\frac{1}{\Lambda}\int \frac{\d^{8|8}X}{\mathrm{vol}\;\GL(2,\C)}\left[(X^2)^{2} \left|\HH^{12}_{12}\right|+\sum_{i,j,k,l}\omega^{1}_{ij}\omega^{2}_{kl}\left|\HH^{12ijkl}_{12ijkl}\right|\right]\;e^{i\cP\cdot X}\;\d^{2}\sigma.
\ee
This expression appears to diverge as $\Lambda\rightarrow 0$, and the leading contribution is a twice-reduced determinant, where each reduction corresponds to the two negative helicity gravitons of the amplitude.  However, the fundamental object in Hodges' formula \eqref{HForm} is a \emph{thrice} reduced determinant; this can be seen as a relic of $\cN=8$ supergravity where all external states are in the same multiplet.  

Despite these apparent roadblocks, we will now demonstrate (building upon calculations which appeared in \cite{Adamo:2012xe}) that $\cM_{n,0}$ is smooth in the $\Lambda\rightarrow 0$ limit, can be given in terms of a thrice-reduced determinant, and reproduces the Hodges' formula in the flat-space limit.

Focusing on the first term in \eqref{3red1}, note that we can represent each factor of $X^{2}$ by a differential `wave operator' acting on $e^{i\cP\cdot X}$:
\be{waveop}
X^{2}\rightarrow \Box :=\frac{I_{IJ}}{(12)}\frac{\partial}{\partial W_{1\;I}}\frac{\partial}{\partial W_{2\;J}}.
\ee
Doing this allows us to re-write the twice-reduced contribution to $\cM_{n,0}$ as
\be{3red2}
\frac{1}{\Lambda}\int \frac{\d^{8|8}X}{\mathrm{vol}\;\GL(2,\C)}\d^{2}\sigma\;\left|\HH^{12}_{12}\right|\;\Box^{2}e^{i\cP\cdot X} =\frac{1}{\Lambda}\int \frac{\d^{2}\sigma}{\mathrm{vol}\;\GL(2,\C)}\left|\HH^{12}_{12}\right|\;\Box^{2}\delta^{8|8}(\cP).
\ee
On the support of this delta-function, we know that the matrix $\HH$ has co-rank three \cite{Hodges:2012ym, Cachazo:2012kg} so we can integrate by parts once with respect to $\frac{\partial}{\partial W_{2}}$ to give
\begin{multline*}
-\frac{1}{\Lambda}\int \frac{\d^{2}\sigma}{\mathrm{vol}\;\GL(2,\C)}\frac{\partial}{\partial W_{2\;J}}\left|\HH^{12}_{12}\right|\frac{I_{IJ}}{(12)}\frac{\partial}{\partial W_{1\;I}}\Box\delta^{8|8}(\cP) \\
=-\int \frac{\d^{2}\sigma}{\mathrm{vol}\;\GL(2,\C)} \sum_{i}\frac{(\xi 2)^{2}}{(12)(i2)(\xi i)^{2}}\left|\HH^{12i}_{12i}\right|\;W_{i}\cdot\frac{\partial}{\partial W_{1}} \Box \delta^{8|8}(\cP).
\end{multline*}
Once again, the support of the delta-function indicates that we can take $W_{i}\cdot\frac{\partial}{\partial W_{1}}=\sigma_{1}\cdot\frac{\partial}{\partial\sigma_{i}}$, and then integrate by parts once again with respect to $\d^{2}\sigma_{i}$.  This leaves us with
\begin{multline}
\int \frac{\d^{2}\sigma}{\mathrm{vol}\;\GL(2,\C)} \sum_{i}\frac{(12)^{2}}{(1i)^{2}(2i)^{2}}\left|\HH^{12i}_{12i}\right|\;\Box\delta^{8|8}(\cP) \\
+\int \frac{\d^{2}\sigma}{\mathrm{vol}\;\GL(2,\C)} \sum_{i,j}\left(\frac{(\xi 2)^{2}(1\xi)(ji)+(\xi 2)^{2}(1j)(\xi i)}{(12)(i2)(ji)(\xi i)(\xi j)^{2}}\right)\HH_{ij}\;\left|\HH^{12ij}_{12ij}\right|\;\Box\delta^{8|8}(\cP).
\end{multline}

The contribution from the second line can be further simplified by noting that the summation entails symmetrization, term-by-term, in both $1\leftrightarrow 2$ and $i\leftrightarrow j$.  A straightforward calculation involving several applications of the Schouten identity allows us to reduce this to
\begin{equation*}
\int \frac{\d^{2}\sigma}{\mathrm{vol}\;\GL(2,\C)} \sum_{i,j}\left(\frac{(\xi 1)^{2}(i2)(j2)+(\xi 2)^{2}(i1)(j1)}{(1i)(2i)(1j)(2j)(\xi i)(\xi j)}\right)\HH_{ij}\;\left|\HH^{12ij}_{12ij}\right|\;\Box\delta^{8|8}(\cP).
\end{equation*}

Upon using the symmetry of $i\leftrightarrow j$ and the basic properties of determinants, we are finally left with an expression for the amplitude with thrice-reduced determinants:
\begin{multline}\label{3red3}
\cM_{n,0}=\int \d^{4|8}x \left[X^{2}\sum_{i,j}\left(\frac{(\xi 1)^{2}(i2)(j2)+(\xi 2)^{2}(i1)(j1)}{(1i)(2i)(1j)(2j)(\xi i)(\xi j)}\right)\left|\HH^{12i}_{12j}\right| \right. \\
\left. X^{2}\sum_{i}\frac{(12)^{2}}{(1i)^{2}(2i)^{2}}\left|\HH^{12i}_{12i}\right| +\frac{1}{\Lambda}\sum_{i,j,k,l}\omega^{1}_{ij}\omega^{2}_{kl}\left|\HH^{12ijkl}_{12ijkl}\right|\right]\;\prod_{m=1}^{n}h_{m}\;\D\sigma_{m},
\end{multline}
where we have reverted to arbitrary twistor wavefunctions.  This shows that the apparent singularity in $\Lambda^{-1}$ of the leading term in \eqref{3red1} is artificial, and also casts $\cM_{n,0}$ in a format based on determinants of $(n-3)\times(n-3)$ matrices.

Now, let us consider the flat-space limit.  Since $\omega^{1,2}_{ij}\sim O(\Lambda)$, it is clear that
\begin{multline}\label{Flat1}
\lim_{\Lambda\rightarrow 0}\cM_{n,0}=\lim_{\Lambda\rightarrow 0}\int \d^{4|8}x \left[X^{2}\sum_{i,j}\left(\frac{(\xi 1)^{2}(i2)(j2)+(\xi 2)^{2}(i1)(j1)}{(1i)(2i)(1j)(2j)(\xi i)(\xi j)}\right)\left|\HH^{12i}_{12j}\right| \right. \\
\left. + X^{2}\sum_{i}\frac{(12)^{2}}{(1i)^{2}(2i)^{2}}\left|\HH^{12i}_{12i}\right|\right]\;\prod_{k=1}^{n}h_{k}\;\D\sigma_{k}
\end{multline}
In the flat-space limit, the second summation in this expression is manifestly independent of $\xi\in\CP^{1}$, and a residue calculation shows that the first summation is also $\xi$-independent (c.f., Lemma 4.4 of \cite{Adamo:2012xe}).  This means that we can set $\xi=\sigma_{1}$ without loss of generality, leaving us with:
\be{Flat2}
\cM_{n,0}(\Lambda=0)=\int \d^{4|8}x\left[\sum_{i,j}\frac{(12)^{2}}{(1i)(1j)(2i)(2j)}\left|\HH^{12i}_{12j}\right| +\sum_{i}\frac{(12)^{2}}{(1i)^{2}(2i)^{2}}\left|\HH^{12i}_{12i}\right|\right]\;\prod_{k=1}^{n}h_{k}\;\D\sigma_{k}.
\ee
In arriving at this expression, we use that $X^{2}\rightarrow 1$ as $\Lambda\rightarrow 0$ and understand that the entries of $\HH$ are computed with respect to a flat-space infinity twistor.

The final step is to realize that on the support of overall momentum conservation, every term in \eqref{Flat2} is equivalent.  This follows from the basic properties of reduced determinants and is built into the Hodges' formula itself, which has many equivalent expressions \cite{Hodges:2012ym, Cachazo:2012kg}.  So up to an irrelevant integer constant (which can be accounted for with proper normalizations), we find:
\begin{equation*}
\lim_{\Lambda\rightarrow 0}\cM_{n,0}=\int \d^{4|8}x\; \frac{(12)^{2}}{(1i)^{2}(2i)^{2}}\left|\HH^{12i}_{12i}\right|\;\prod_{j=1}^{n}h_{j}\;\D\sigma_{j}= \cM^{\mathrm{Hodges}}_{n,0}(\Lambda=0),
\end{equation*}
as required.

\section{Conclusions and further directions}
\label{DC}

In this paper, we have developed the perturbative analysis of the twistor action for conformal gravity so as to obtain the MHV amplitudes.  We then went on to use the embedding of Einstein states into conformal gravity to deduce new formulae for MHV amplitudes in de Sitter (and AdS) for Einstein gravity.  An off-shell version of this embedding led to the proposed new action functionals for Einstein gravity itself.

The conformal gravity twistor action is classically equivalent off-shell to the space-time action following from theorem \ref{MThm}.   What we do not currently have is an analogue of theorem \ref{MThm} for Einstein gravity, largely because at this stage the geometric structure in the Einstein case is presented in a coordinate form.  This is clearly a major outstanding question which we hope to address in future work.  Although a proposal was made for an Einstein twistor action with $\Lambda=0$ in \cite{Mason:2008jy}, this is more robust because in that case the MHV generating formula was only ever presented in a gauge fixed form, whereas here the gauge is not so completely fixed and there is the possibility of transforming to a space-time gauge.  Our Einstein twistor action \eqref{EinTA4} has the correct self-dual reduction, produces the correct MHV amplitude (i.e., the same as obtained via conformal gravity), and is derived from the Einstein gravity embedding in conformal gravity.  The Einstein twistor action (if correct) would be an important tool as, in principle, it would allow us to perform loop computations. 

The new formulae are striking in view of their structure as rank $n-2$ determinants that degenerate as $\Lambda\rightarrow 0$, rather than the rank $n-3$ generalized determinant of Hodges.  This is perhaps reminiscent of the two versions of the KLT relations \cite{BjerrumBohr:2010ta}.  
We discuss further issues in separate sections below.


\subsection{Axial gauge and the MHV formalism}

One of the important applications of the twistor action for $\cN=4$ SYM is that it leads to a derivation of the MHV formalism \cite{Cachazo:2004kj} for Yang-Mills by virtue of an axial gauge choice \cite{Boels:2007qn, Adamo:2011cb}.  The key benefit of this gauge choice is that it exploits the integrability of the self-dual sector, essentially trivializing it by knocking out the non-linear terms in the self-dual part of the action so that the only vertices are those arising from the non-local part.  The existence of a MHV formalism for gravity remains controversial \cite{BjerrumBohr:2005jr, Bianchi:2008pu}.\footnote{It is worth mentioning the recent work of \cite{Penante:2012wd}, which proposes a MHV-like formalism based on delta-function relaxation in a Grassmannian representation of the gravitational amplitudes \cite{He:2012er, Cachazo:2012pz}.}   Nevertheless, we saw at the beginning of Section \ref{MACC} that the axial gauge can also be imposed on the twistor actions for conformal and Einstein gravity, resulting in a twistor space propagator $\Delta(Z,Z')$ and Riemann sphere propagator $\Delta(\sigma,\sigma')$. 

The propagator $\Delta(Z,Z')$ was not used in this paper, but it will essentially be the same as in the Yang-Mills case but with different weights to account for the different helicities involved.  We refer the reader to \cite{Mason:2010yk, Adamo:2011cb, Adamo:2011pv} for the definition of this propagator in the Yang-Mills case; it is essentially a delta function constraining $Z, Z'$ and $Z_*$ to be collinear in twistor space, and with a Cauchy pole when $Z$ and $Z'$ come together.  The effect of the Riemann sphere propagator and the $Y\cdot f$ vertices is wrapped up into the full MHV amplitude.  This can still be extended off-shell to become a MHV \emph{vertex} on twistor space by allowing ourselves to insert one or other end of the propagator  $\Delta(Z,Z')$ instead of the $f$ or $g$ (or $h$ and $\tilde h$ in the Einstein case).  Thus the propagator connects the vertices which, on-shell, give the MHV amplitude, so it induces a MHV formalism for Einstein gravity in twistor space.  

To compute a N$^k$MHV amplitude in conformal gravity reduced to Einstein gravity, we must sum diagrams with $k+1$ MHV vertices and $k$ propagators, then divide by the overall factor of $\Lambda$ required by the embedding of Einstein gravity in conformal gravity.  If our proposed Einstein twistor action is correct, this should give the same answer as the computation using $k$ propagators from the Einstein gravity twistor action replacing $h$ and $\tilde h$ in the vertices.  Compatibility would essentially provide a proof that our proposal is correct (at least at the level of perturbation theory).  Preliminary calculations indicate that they are in fact compatible and we hope to pursue this elsewhere.

The framework developed in this paper is sufficient for computing formulae for gravity amplitudes in twistor space along the lines of \cite{Adamo:2011cb}; there is much work needed to make contact with momentum space formulae, though.   It is interesting to note that the structure of the twistorial MHV formalism differs significantly from what has been proposed previously in momentum space formulae.  In particular, the functional form of $\cM_{n,0}$ begins with a twice-reduced determinant, as in \eqref{MHVamp} or \eqref{MHVamp2}.  While we were able to obtain a thrice-reduced determinant form in \eqref{3red3}, the arguments which produced this were on-shell in nature, making them unsuitable for treating a vertex instead of an amplitude.  This indicates that in flat space, the MHV formalism that arises from this twistor action will \emph{not} simply correspond to an off-shell extension of the Hodges formula linked with $p^{-2}$ propagators; or at least, not in any obvious way.  Rather, we might expect an off-shell extension of the twice-reduced formula \eqref{MHVamp}, with a propagator prescription given by translating $\Delta(Z,Z')$ to momentum space.


\subsection{Connections to the $\cN=8$ twistor-string formulae}

Equation \eqref{3red3} is most closely related to the formulae that arise from Skinner's $\cN=8$ twistor-string, which makes direct contact with the Hodges formula at MHV.  Skinner's $\cN=8$ twistor-string is the first example of a theory which treats Einstein supergravity \emph{directly} with twistor methods \cite{Skinner:2013xp}.  As a string theory, it is anomaly free for any genus worldsheet and is known to produce the complete tree-level S-matrix of $\cN=8$ supergravity on a flat background.  Furthermore, the worldsheet theory is perfectly well-defined for a non-simple infinity twistor, so in principle it should also be able to produce (after truncation to $\cN=4$ supersymmetry) the same twistor space formulae we have derived here.

Unfortunately, it is not currently known how to compute meaningful worldsheet correlators of gravitational vertex operators with a cosmological constant in Skinner's twistor-string (beyond three-points).  The issues which arise are the failure of the correlators to be independent of the position of picture changing operators as well as reference spinors (analogous to $\xi\in\CP^1$); this indicates that the correlators are not gauge invariant with respect to the worldsheet degrees of freedom.  These problems could stem from any number of sources, including an incomplete understanding of the full spectrum of vertex operators for the theory, or the worldsheet Feynman rules when $\Lambda\neq 0$.  Hence, it seems natural to ask if our formulae for $\cM_{n,0}$ could shed any light on this twistor-string calculation.

Just as in Hodges formula, the fundamental object for computing the MHV amplitude in Skinner's twistor-string is a thrice-reduced determinant, corresponding to building a top-degree form on the space of fermionic automorphisms of the worldsheet \cite{Skinner:2013xp}.   Note that not only does \eqref{3red3} have the desired thrice-reduced determinants, but it also features Vandermonde factors in the coordinates $\sigma_{i}$ which arise in the context of twistor-string theory.   In particular, it is not immediately clear how the final term (with a six-times reduced determinant) might arise from the twistor-string theory and this might provide some clues as to how to better understand computations in that theory at $\Lambda\neq 0$.


\subsection{Physical observables}

Throughout this paper, we have referred to our formula for the MHV amplitude $\cM_{n,0}$ as a `scattering amplitude' for general relativity on a background with cosmological constant.  As pointed out in the introduction, we have adopted this terminology for convenience and the notion of a physically observable scattering amplitude on de Sitter space is not completely well-defined.  The final formulae we obtain for $\cM_{n,0}$  on twistor space make mathematical sense for arbitrary choices of twistor wavefunctions $h_i$ and $\tilde h_i$, as well as integration region or contour for the $\d^{4}x$ (or $\d^{4|8}x$ in the $\cN=4$ case) integral in conformally compactified Minkowski space.  \emph{A priori}, the twistor wavefunctions can be the Penrose transform of choice of space-time wave-function, although for convenience we used the dual `elemental' twistor states \eqref{dtwf} for calculational purposes in Section \ref{MACC}.  More usually, amplitudes are expressed in terms of momentum eigenstates.  Twistor wavefunctions that correspond to \emph{momentum eigenstates} (c.f., \cite{Adamo:2011pv}), with four-momentum $k_{AA'}=p_{A}\tilde{p}_{A'}$ are given by
\be{momeig}
h(Z(\sigma), k_{AA'})=\int_{\C}\frac{\d s}{s^{1+w}}\bar{\delta}(s\lambda_{A}-p_{A})\;\e^{s[\mu\tilde{p}]},
\ee
where $w=-6$ for a negative helicity graviton and $w=2$ for a positive helicity graviton.

These are somewhat unnatural from the point of view of de Sitter geometry, since there is no four dimensional abelian subgroup for the de Sitter group.  They nevertheless make sense conformally and can be set up with respect to either of the flat coordinate backgrounds in the coordinate forms in \eqref{dSmetric2} or \eqref{dSmetric3}.  In the first of these, the infinity of the coordinate patch is the lightcone of a finite point and so these eigenstates are singular on this finite light cone and don't recognize the infinity of global de Sitter space. This is nevertheless the more convenient representation for studying the $\Lambda\rightarrow 0$ limit.

The second coordinate system is more satisfactory physically because the infinity of the coordinate system is now the lightcone of a choice of a point at infinity.   Then one can consider the half of the space-time to the past of that light cone to be the observable universe of a physical observer.  Furthermore, at least the 3-dimensional abelian subgroup of spatial translations is a subgroup of the de Sitter group, so these make more sense in this context.

Similarly, we need to choose a contour for the integral over $\d^{4}x$ which corresponds to the real slice of space-time.  Integrating over the full real slice of de Sitter space is equivalent to integrating over the full conformal compactification and so doesn't require any choice of coordinates.  It corresponds to computing a $\scri^{-}$ to $\scri^{+}$ scattering process where $\scri^\pm$ are the future/past space-like infinities of de Sitter space.  Although no observer could measure this, the theory knows how to compute this amplitude and so this has sometimes become known as a meta-observable \cite{Witten:2001kn, Strominger:2001pn}.  Using the eigenstates \eqref{momeig} and in the affine patch \eqref{dSmetric2}, the three-point amplitude in this set-up is \cite{Adamo:2012nn}
\be{3ptMHV}
\cM_{3,0}=\frac{\la 12\ra^{6}}{\la 23\ra^{2} \la 31\ra^{2}}\left(2-\Lambda\Box_{k}\right)\delta^{4}\left(\sum_{i=1}^{3}k_{i}\right),
\ee
where $\Box_{k}$ is the wave operator on momentum space.  It limits onto the definition of the scattering amplitude when $\Lambda\rightarrow 0$.  Using this prescription for $\cM_{n,0}$ will produce an operator of leading order $\Box_{k}^{n-2}$.

To obtain an honest physical observable, one should use twistor momentum eigenstates \eqref{momeig} but now adapted to de Sitter space in the form \eqref{dSmetric3}, and choose the contour of integration for $\d^{4}x$ to correspond to a physically observable region of $dS_{4}$, say $t>0$ and the integration contour can be displaced into the complex so as to coincide with the in-in formalism.  Prescriptions of this sort have been used to calculate the non-Gaussianities in the gravitational bispectrum from inflation (c.f., \cite{Maldacena:2002vr, Maldacena:2011nz}).  


\subsection{Space-time background-coupled calculations}

A challenge for these kinds of techniques is to extend the calculation to one off other backgrounds; particularly interesting choices might be black holes or even plane waves.  Indeed, the calculation here can already be viewed in these terms, but as one off a self-dual background as in Appendix \ref{GenFuncs}.  Restricting to the axial gauge on twistor space and considering only MHV amplitudes removes the twistor propagator $\Delta(Z,Z')$ from the calculations.  In the Einstein case, the remaining elements of perturbation theory were the vertices $V_{h}$, $V_{\tilde{h}}$, and the propagator $\Delta(\sigma,\sigma')$.  To compute the $n$-point MHV amplitude, the addition of the $n-2$ $V_{h}$ vertices is equivalent to expanding to $(n-2)^{\mathrm{th}}$ order the vertex: 
\be{Vertgen}
\frac{1}{\Lambda}\int_{M}\d^{4|8}x \left(\int_{X} \tilde{h}\wedge\tau\right)^2\, ,
\ee
which is evaluated on the fully non-linear self-dual background space-time $M$.

By theorem \ref{NLG}, $M$ can be obtained via the non-linear graviton construction.  The equation for the holomorphic curve $X$ now reduces to
\be{holomap2}
\dbar_{\sigma} Z^{I}(x,\sigma)=I^{IJ}\partial_J h(Z)\, ,
\ee
which has the four complex parameter family of solutions defining (complexified) space-time \cite{Penrose:1976js, Hansen:1978jz}.\footnote{When $\Lambda=0$, it is a twistorial formulation of the `good cut equation' \cite{Eastwood:1982, Adamo:2010ey, Adamo:2009vu}.}  The Feynman diagram formalism on $\CP^1$ introduced in Section \ref{MACC} generates the perturbative solution to \eqref{holomap2}, substituted into \eqref{Vertgen}.  Since the curve $X\subset\CPT$ is constructed from the classical solution to \eqref{holomap2}, only tree diagrams contribute to the Feynman diagram calculus.  This provides a geometric viewpoint for the Feynman rules of the twistor action which live on the Riemann sphere: they operationalize the perturbative expansion of the non-linear self-dual background $M$.  

It should therefore be possible to understand $\cM_{n,0}$ in purely geometric terms, without reference to a Feynman diagram formalism.  That is, one should be able to derive our tree formulae by iteratively solving \eqref{holomap2} around the background built from the $n-2$ positive helicity scattering states.  It would be interesting to see if backgrounds more complicated than the self-dual one studied here can also be described in this way.

\acknowledgments

We thank Eduardo Casali, Yvonne Geyer, and especially David Skinner for many useful conversations.  TA is supported by a National Science Foundation (USA) Graduate Research Fellowship and has benefited from the hospitality of the I.H.\'E.S. during the completion of this work; LM is supported by a Leverhulme Fellowship and EPSRC grant number EP/J019518/1.

\appendix


\section{Appendices}
\label{Appends}

\subsection{The conformal geometry of de Sitter space}\label{desitter-geom}

De Sitter, anti-de Sitter, and flat space-times in $n$-dimensions possess only scalar curvature and are hence conformally flat.  Each is a dense open subset in the conformal compactification which is a projective quadric of signature $(2,n)$ in $\RP^{n+1}$ of topology $S^1\times S^{n-1}/\Z_2$.  The infinite points are respectively a space-like, time-like or null hypersurface (in fact a lightcone) in the conformal compactification obtained as the intersection of a hyperplane of appropriate signature in $\RP^{n+1}$.   

\begin{figure}
\centering
\includegraphics[width=2 in, height=1.7 in]{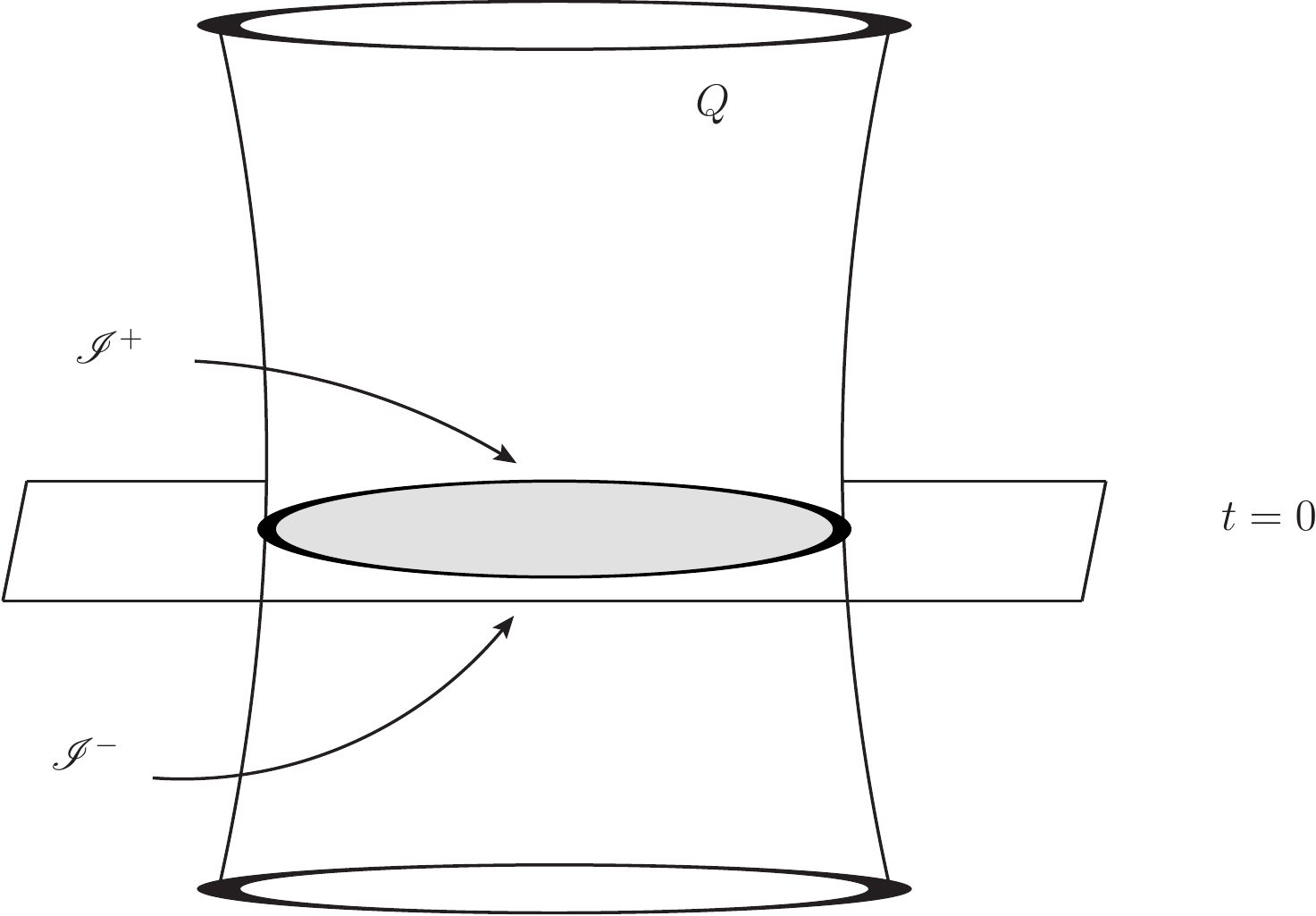}\caption{\textit{De Sitter space as the quadric} $Q\subset\RP^{5}$ \textit{and the identification of infinity.}}\label{dS1}
\end{figure}

In four dimensions, de Sitter space ($dS_{4}$) is topologically $\R\times S^{3}$, and can be realized as the pseudosphere in $\R^{1,4}$ with coordinates $(w, x^{\mu})$, $\mu=0,\ldots, 3$ via the embedding \cite{Hawking:1973}:
\begin{equation*}
\eta_{\mu\nu}x^{\mu}x^{\nu}-w^{2}=x^{2}-w^{2}=-\frac{3}{\Lambda}, \qquad \eta_{\mu\nu}=\mathrm{diag}(1,-1,-1,-1)\, .
\end{equation*}
This makes manifest the isometry group $\SO(1,4)$, the Lorentz group inherited from the embedding space.  The embedding as a projective quadric in $\RP^{5}$ can be realized with homogeneous coordinates $(t,w,x^{\mu})$ as the $t\neq 0$ portion of:
\begin{equation*}
2Q\equiv t^{2}-w^{2}+x^{2}=0,
\end{equation*}
with scale-invariant metric
\be{dSmetric1}
\d s^{2}=\frac{3}{\Lambda}\frac{\d t^{2}-\d w^{2}+\eta_{\mu\nu}\d x^{\mu}\d x^{\nu}}{t^{2}}.
\ee
The intersection of $Q$ with the plane $t=0$ corresponds to the spatial $S^{3}$ at infinity, and is the identification of the past ($\scri^{-}$) and future ($\scri^{+}$) infinities (ordinarily, we will not make this identification); see Figure \ref{dS1}.  The pseudosphere in $\R^{1,4}$ is recovered by taking the patch $t=\sqrt{3/\Lambda}$.

\begin{figure}
\centering
\includegraphics[width=3.25 in, height=1.5 in]{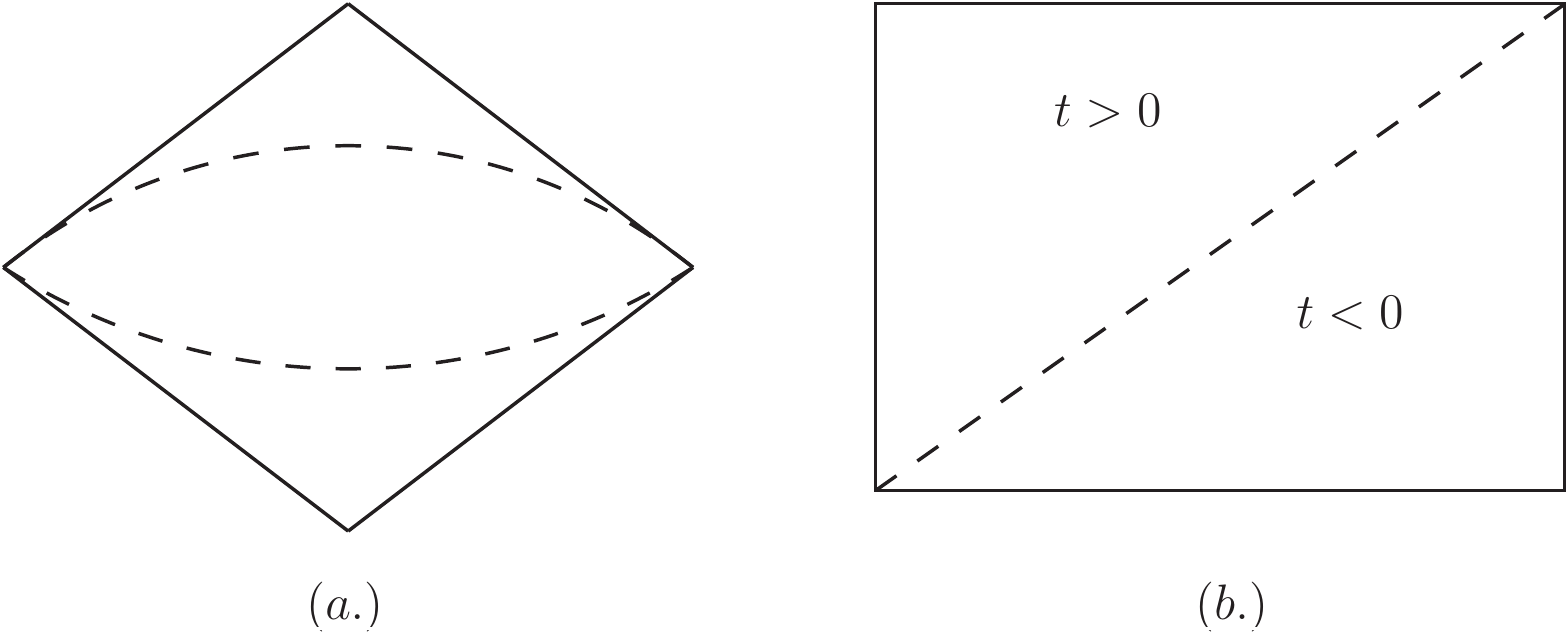}\caption{\textit{De Sitter space on the affine Minkowski patch} (\emph{a}.), \textit{and the Poincar\'{e} patch} (\emph{b}.)}\label{dS2}
\end{figure}

On de Sitter space, one can work with various convenient coordinate patches. We note here two conformally flat  choices.  The first and most commonly used one is the Poincar\'e patch  which corresponds to $x^{0}+w=1$, with metric:
\be{dSmetric3}
\d s^{2}=\frac{3}{\Lambda}\frac{\d t^{2}-\delta_{ij}\d x^{i}\d x^{j}}{t^{2}}.
\ee
The $t=0$ slice is infinity minus a point whose light cone divides de Sitter space into two halves ($t>0$ and $t<0$), demonstrating that a physical observer at $\scri^{\pm}$ has access to at most half of the space-time.  The Poincar\'e patch manifests the three-dimensional rotation and translation symmetries of $dS_{4}$, but is not so well-behaved in the $\Lambda\rightarrow 0$ limit; see Figure \ref{dS2}, (\emph{b}.).

A less conventional choice is an affine patch in which the vertex of the light cone at infinity for the affine coordinate patch is taken to be at a finite point.  This corresponds to $t+w=1$ and after re-scaling the affine Minkowski coordinates $x^{\mu}$ the metric becomes
\be{dSmetric2}
\d s^{2}=\frac{\eta_{\mu\nu}\d x^{\mu} \d x^{\nu}}{(1-\Lambda x^{2})^2}.
\ee
Most of de Sitter infinity is then located at finite points in the affine space where $x^{2}=\Lambda^{-1}$, although this has an $S^2$ intersection with the affine coordinates' infinity.  Here the flat space-time emerges as $\Lambda\rightarrow 0$; see Figure \ref{dS2}, (\emph{a}.).  


\subsection{The MHV amplitude as a SD background coupled field calculation}
\label{GenFuncs}


We will focus on the tree-level Einstein Maximal Helicity Violating (MHV) amplitudes which correspond to the scattering of two negative helicity gravitons and $n-2$ positive helicity gravitons.  These are maximal because the positive and negative helicity states are dual to each other, so an `all +' amplitude would correspond to a positive helicity particle picking up some negative helicity scattering on a positive helicity background.  But this cannot happen by virtue of the consistency of the self-duality equations for general relativity.  Similarly, the one negative and rest positive helicity amplitude vanishes because the self-dual sector is integrable (it would correspond to the non-trivial scattering of a linear positive helicity particle on a positive helicity background).  See lemma \ref{SDsec} below for more details.  

\begin{figure}
\centering
\includegraphics[width=3.6 in, height=1.5 in]{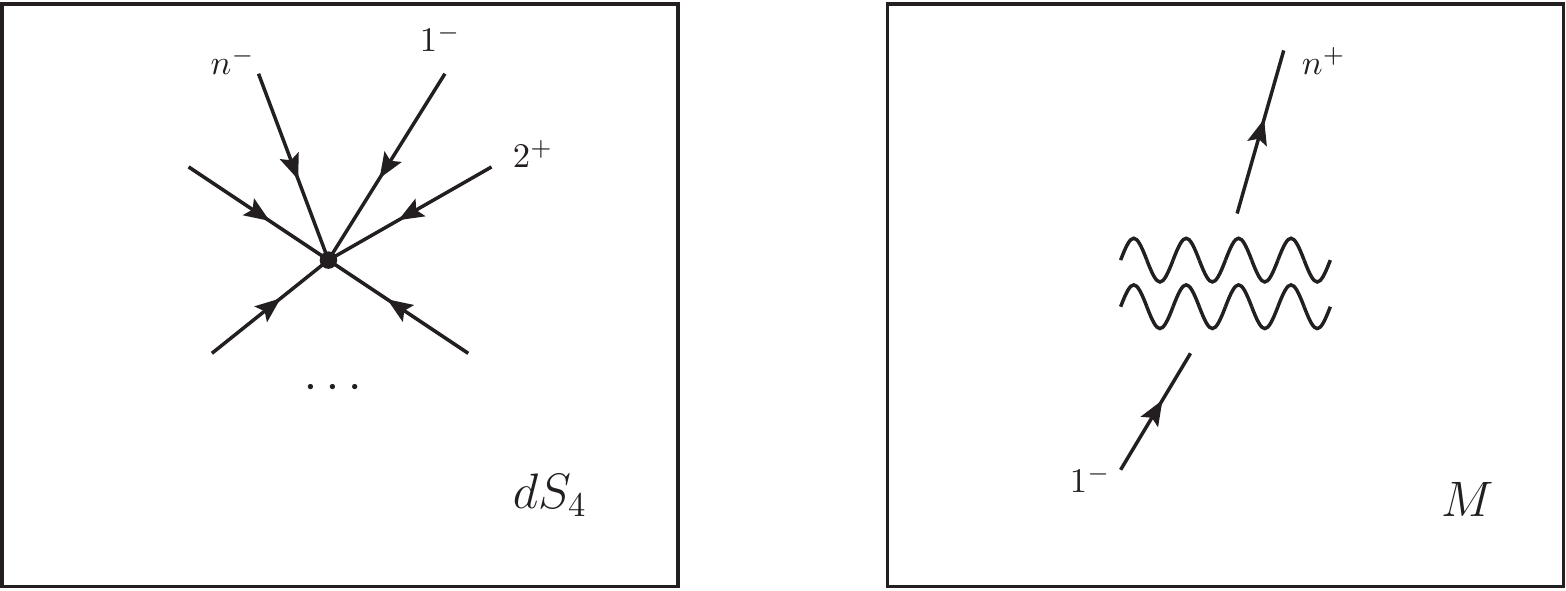}\caption{\textit{Geometric picture of MHV graviton scattering}}\label{dS3}
\end{figure}

Following \cite{Mason:2008jy}, we absorb the $n-2$ SD gravitons of the MHV amplitude into a fully nonlinear SD background space-time $M$, which can subsequently be perturbatively expanded to recover the individual particle content.  Reversing the momentum of one of the two negative helicity gravitons, the MHV amplitude is the probability for a pure ASD state at $\scri^{-}$ to propagate across $M$ and evolve into a SD state at $\scri^{+}$ as illustrated in Figure \ref{dS3}.

The generating functional for MHV amplitudes in \emph{conformal gravity} is given by the second term in \eqref{CGA3}.  The first term is precisely the action for the self-dual sector, so the second term is therefore the action for the first non-trivial deformation of the SD sector that is quadratic in the ASD part of the field.  Evaluated on-shell with Einstein scattering states, the two ASD gravitons are given by Weyl spinor perturbations $\psi_{1}$, $\psi_{2}$ and the generating functional reads:
\be{CGGF}
\left. I^{\mathrm{CG}}[1^{-}, 2^{-}; M^{+}]\right|_{\mathrm{Ein}}=\frac{2i}{\varepsilon^{2}}\int_{M} \d\mu \; \psi_{1}^{ABCD}\psi_{2\;ABCD},
\ee
where $M$ is again the SD background which encodes the $n-2$ remaining gravitons.  

We now derive the generating functional for these amplitudes in Einstein gravity using the Einstein-Plebanski action to perturb about the SD sector in proposition \ref{MHVGF}. This leads to a generating functional that we denote by $I^{\mathrm{GR}}$ and is given in \eqref{MHV1}.  In proposition \ref{CGDS}, we show that on a self-dual background this arises from the reduction of \eqref{CGGF} to Einstein data, so that
\be{CGDS*}
I^{\mathrm{GR}}[1^{-}, 2^{-}; M^{+}]=-\frac{3 \varepsilon^{2}}{\Lambda\;\kappa^{2}}\left. I^{\mathrm{CG}}[1^{-}, 2^{-}; M^{+}]\right|_{\mathrm{Ein}},
\ee
in accordance with \eqref{CGB2}.  Thus we can use either as a generating function to compute the MHV amplitude.  

We will exploit the chiral formulation of general relativity \cite{Plebanski:1977zz}: for a general space-time $M$ with metric specified by a tetrad of 1-forms $\d s^{2}=\epsilon_{AB}\epsilon_{A'B'}e^{AA'}\otimes e^{BB'}$, the basic variables are three ASD 2-forms:
\begin{equation*}
\Sigma^{AB}=e^{A'(A}\wedge e^{B)}_{A'},
\end{equation*}
and the ASD spin connection $\Gamma_{AB}$. With a cosmological constant $\Lambda$, the action of general relativity is:
\be{eqn: PA}
S[\Sigma, \Gamma]=\frac{1}{\kappa^2}\int_{M} \left(\Sigma^{AB}\wedge F_{AB}-\frac{\Lambda}{6}\Sigma^{AB}\wedge\Sigma_{AB}\right),
\ee
where 
\be{eqn: ASDcurv}
F_{AB}=\d\Gamma_{AB}+\Gamma^{C}_{A}\wedge \Gamma_{BC}
\ee
is the curvature of the ASD spin connection.  This action produces two field equations, to which we append a third (the condition that $\Sigma^{AB}$ be derived from a tetrad) \cite{Capovilla:1991qb}:
\begin{eqnarray}
\D \Sigma^{AB} & = & 0, \label{FE1} \\
F_{AB} & = & \Psi_{ABCD}\Sigma^{CD}+\frac{\Lambda}{3}\Sigma_{AB}, \label{FE2} \\
\Sigma^{(AB}\wedge\Sigma^{CD)} & = & 0 . \label{FE3}
\end{eqnarray}
Here, $\D$ is the covariant derivative with respect to the ASD spin connection:
\begin{equation*}
\D\Sigma^{AB}=\d\Sigma^{AB}+2\Gamma^{(A}_{C}\wedge\Sigma^{B)C}.
\end{equation*}

Following \cite{Mason:2008jy}, we can express a tree-level MHV amplitude as the classical scattering of two negative helicity gravitons off a SD background space-time, which (perturbatively) encodes the remaining positive helicity gravitons.  For a SD background, we have $\Psi_{ABCD}=0$, so \eqref{FE2} can be solved for $\Sigma$ in terms of $F$ while \eqref{FE1}, \eqref{FE3} result in an algebraic condition on the curvature of the ASD spin connection.  To be precise, a SD solution $(\Sigma_{0}, \Gamma_{0})$ obeys \cite{Capovilla:1990qi}:
\begin{eqnarray}
\Sigma_{0}^{AB} & = & \frac{3}{\Lambda} F^{AB}_{0}, \label{SD1} \\
F_{0(AB}\wedge F_{0\; CD)} & = & 0. \label{SD2}
\end{eqnarray}

Now consider small perturbations away from this SD background of the form $\Sigma= \Sigma_{0}+\sigma_{0}$, $\Gamma = \Gamma_{0}+\gamma$.  This results in a set of linearized field equations:
\begin{eqnarray}
\D_{0}\sigma^{AB} & = & -2\gamma^{(A}_{C}\wedge\Sigma^{B)C}_{0}, \label{LFE1} \\
\D_{0}\gamma_{AB} & = & \psi_{ABCD}\Sigma^{CD}_{0}+\frac{\Lambda}{3}\sigma_{AB}, \label{LFE2} \\
\sigma^{(AB}\wedge\Sigma^{CD)}_{0} & = & 0, \label{LFE3}
\end{eqnarray}
where $\D_{0}$ is the covariant derivative with respect to the background ASD spin connection $\Gamma_{0}$.  It is fairly easy to see that the field $\psi_{ABCD}$ corresponds to a linearized ASD Weyl spinor propagating on the SD background $(\Sigma_{0},\Gamma_{0})$ \cite{Adamo:2013}.

Our goal is now to formalize the picture of an MHV amplitude in terms of linearized solutions propagating on a SD background.  If $\mathcal{S}$ is the space of solutions to the full field equations \eqref{FE1}-\eqref{FE3}, then solutions to the linearized equations \eqref{LFE1}-\eqref{LFE3} are a vector space $V$ corresponding to the fiber of $T\mathcal{S}$ over the SD solution $(\Sigma_{0},\Gamma_{0})$.  Now, a linearized SD solution is fully characterized by the ASD spin connection, since
\be{eqn: SDLFE}
\sigma_{AB} =\frac{3}{\Lambda}\D_{0}\gamma_{AB}, \qquad \D_{0}\gamma^{(AB}\wedge F_{0}^{CD)} = 0.
\ee
This allows us to define the SD portion of $V$ as
\begin{equation*}
V^{+}=\left\{(\sigma,\gamma)\in V \: : \: \D_{0}\gamma^{(AB}\wedge F_{0}^{CD)} = 0\right\},
\end{equation*}
and a corresponding $V^{-}$ by the quotient map in the short exact sequence:
\begin{equation*}
0\longrightarrow V^{+} \hookrightarrow V \longrightarrow V^{-} \longrightarrow 0.
\end{equation*}
In particular, this means we have
\begin{equation*}
V^{-}\equiv V/ V^{+}= \left\{(\sigma,\gamma)\in V\right\} / \left\{\gamma \: : \: \D_{0}\gamma^{(AB}\wedge F_{0}^{CD)} = 0\right\} .
\end{equation*}

The space of solutions $\mathcal{S}$ comes equipped with a natural symplectic form $\omega$ given by the boundary term in the action \cite{Ashtekar:2008jw}:
\be{eqn: symp}
\omega = \frac{1}{\kappa^{2}}\int_{C}\delta\Sigma^{AB}\wedge\delta\Gamma_{AB},
\ee
where $C$ is a Cauchy surface in $M$ (when $\Lambda>0$, there is always a slicing where $C\cong S^3$ topologically) and $\delta$ is the exterior derivative on $\mathcal{S}$.  It is straightforward to show that $\omega$ is independent of the choice of Cauchy surface and descends to a symplectic form on $\mathcal{S}/\mathrm{Diff}^{+}_{0}(M)$ \cite{Adamo:2013}.

This symplectic form induces an inner product between points in the linearized solution space $V$: for $h_{i}, h_{j}\in V$ we take
\be{ip}
\la h_{i}|h_{j}\ra = -\frac{i}{\kappa^{2}}\int_{C}\sigma^{AB}_{j}\wedge\gamma_{i\;AB}.
\ee
An important fact about this inner product (which is obvious in the $\Lambda=0$ setting, c.f., \cite{Mason:2008jy}) is that it annihilates the SD sector:
\begin{lemma}\label{SDsec}
Let $h_{i},h_{j}\in V^{+}$ on the SD background with $(\Sigma_{0},\Gamma_{0})$.  Then $\la h_{i}|h_{j} \ra=0$, or equivalently: for all $h_{i}\in V^{+}$, $\la h_{i}|\cdot\ra|_{V^+}=0$.
\end{lemma}
\proof  The inner product is skew-symmetric under interchange of $h_{i}$ and $h_{j}$, so
\begin{equation*}
\la h_{i}|h_{j} \ra =-\frac{i}{2\kappa^{2}}\int_{C}\left(\sigma^{AB}_{j}\wedge\gamma_{i\;AB}-\sigma^{AB}_{i}\wedge\gamma_{j\;AB}\right).
\end{equation*}
Suppose $h_{j}\in V^{+}$; then \eqref{LFE2} implies that $\D_{0}\gamma_{j\;AB}=\frac{\Lambda}{3}\sigma_{j\;AB}$.  In the $\Lambda=0$ limit, the ASD spin connection is trivial $\D_{0}\rightarrow\d$, so $\gamma_{j}^{AB}|_{\Lambda=0}=0$, and we can write $\gamma_{j}^{AB}=\Lambda \nu_{j}^{AB}$ for some array of space-time 1-forms $\nu_{i}^{AB}$.  With this representation, the linearized SD field equation gives $\sigma_{j\;AB}=3\D_{0}\nu_{j\;AB}$, and the inner product becomes:
\begin{multline*}
-\frac{i}{2\kappa^{2}}\int_{C}\left(3\d\nu^{AB}_{j}\wedge\gamma_{i\;AB}+6\Gamma_{0\;C}^{(A}\wedge\nu^{B)C}_{j}\wedge\gamma_{i\;AB}-\sigma^{AB}_{i} \wedge\gamma_{j\;AB}\right)\\
=\frac{i}{2\kappa^{2}}\int_{C}\left(3\nu^{AB}_{j}\wedge\D_{0}\gamma_{i\;AB}-\sigma^{AB}_{i}\wedge\gamma_{j\;AB}\right),
\end{multline*}  
where the second line follows by integration by parts and a re-arranging of index contractions.  Once again using $\gamma_{j\;AB}=\Lambda\nu_{j\;AB}$, we have:
\begin{equation*}
\la h_{i}|h_{j}\ra = \frac{i}{2\kappa^{2}}\int_{C}\nu_{j}^{AB}\wedge\left(3\D_{0}\gamma_{i\;AB}-\Lambda\sigma_{i\;AB}\right)=\frac{3i}{2\kappa^{2}}\int_{C}\nu_{j}^{AB}\wedge\psi_{i\;ABCD}\Sigma^{CD}_{0},
\end{equation*} 
using \eqref{LFE2} for $h_{i}$.  Hence, if $h_{i}\in V^{+}$ then $\psi_{i\;ABCD}=0$ and the inner product vanishes.     $\Box$

\medskip

Note that lemma \ref{SDsec} confirms that the all-positive helicity and $(-+\cdots +)$ amplitudes of general relativity vanish even with a cosmological constant in play.  In the first case, we see that the SD field equations are integrable since their solutions are characterized by a single algebraic relation \eqref{SD2}.  In the second case, the fact that the inner product annihilates the SD sector ensures that scattering with only a single negative helicity graviton is also trivial.

We can use this inner product to define ASD solutions at the boundary of our $M$ as in \cite{Mason:2008jy}: take a one-parameter family of Cauchy hypersurfaces $C_{t}\rightarrow\scri^{\pm}$ as $t\rightarrow\pm\infty$.  Then we say that $h_{j}=(\sigma_{j},\gamma_{j})$ is ASD at $\scri^{\pm}$ if
\be{ASDlim}
\lim_{t\rightarrow\pm\infty} \int_{C_t}\sigma^{AB}_{j}\wedge\gamma_{i\;AB}=0 \qquad \mbox{for all}\:\: h_{i}=(\sigma_{i},\gamma_{i})\in V^{-}.
\ee

Now we want to build the generating functional for the MHV amplitudes, which measure the probability for a pure ASD state at $\scri^{-}$ to propagate across a SD background $M$ and evolve into a SD state at $\scri^{+}$.  Hence, we take the incoming state to be $h_{1}|_{\scri^{-}}\in V^{-}$.  Since the inner product annihilates the SD sector, we need to compute the inner product between $h_{1}$ and some other state $h_{2}|_{\scri^{+}}\in V^{-}$ at the future conformal boundary $\scri^{+}$:\footnote{As mentioned in the text, this corresponds to a `meta-observable' since we integrate over the entire space-like surface $\scri^{+}$.}  This gives the generating functional for the MHV amplitudes as
\be{ip*}
I^{\mathrm{GR}}[1^{-},2^{-},M^{+}]=\la h_{2}|h_{1}\ra =-\frac{i}{\kappa^{2}}\int_{\scri^{+}}\sigma^{AB}_{1}\wedge\gamma_{2\;AB}.
\ee
This form of the generating functional is not particularly illuminating because the role of the SD background $M$ is implicit.  However, we can manipulate \eqref{ip*} into a format which is explicitly in terms of an integral over the entire background space-time.

\begin{propn}\label{MHVGF}
The amplitude $\la h_{n}|h_{1}\ra$ is given by the formula:
\be{MHV1}
I^{\mathrm{GR}}[1^{-},2^{-},M^{+}] =\frac{i}{\kappa^{2}}\int_{M}\left(\Sigma^{AB}_{0}\wedge\gamma_{1\;A}^{C}\wedge\gamma_{2\;CB}-\frac{\Lambda}{3}\sigma^{AB}_{1}\wedge\sigma_{2\;AB}\right),
\ee
where $M$ is a SD background space-time described by $(\Sigma_{0},\Gamma_{0})$.
\end{propn}
\proof  Recall that $\partial M=\scri^{+}-\scri^{-}$, so Stokes' theorem gives
\begin{equation*}
-\frac{i}{\kappa^{2}}\int_{\scri^{+}}\sigma^{AB}_{1}\wedge\gamma_{2\;AB}=-\frac{i}{\kappa^{2}}\int_{M}\left(\d\sigma_{1}^{AB}\wedge\gamma_{2\;AB}+\sigma^{AB}_{1}\wedge\d\gamma_{2\;AB}\right)-\frac{i}{\kappa^{2}} \int_{\scri^{-}}\sigma^{AB}_{1}\wedge\gamma_{2\;AB}.
\end{equation*}
Now, the second term on the right vanishes, since $h_{1}\in V^{-}$ at $\scri^{-}$.  Using the linearized field equations \eqref{LFE1}, \eqref{LFE2} it follows that
\begin{eqnarray*}
\d\sigma_{1}^{AB} & = & -2\gamma_{1\;C}^{(A}\wedge\Sigma^{B)C}_{0}-2\Gamma_{0\;C}^{(A}\wedge\sigma_{1}^{B)C},\\
\d\gamma_{2\;AB} & = & \psi_{2\;ABCD}\Sigma^{CD}_{0}+\frac{\Lambda}{3}\sigma_{2\;AB}-2\Gamma_{0\;C(A}\wedge\gamma_{2\;B)}^{C},
\end{eqnarray*}
and the generating functional becomes
\begin{multline*}
\frac{i}{\kappa^{2}}\int_{M}\left(\Sigma_{0}^{AB}\wedge\gamma_{1\;A}^{C}\wedge\gamma_{2\;CB}+\sigma^{AB}_{1}\wedge\Gamma_{0\;A}^{C}\wedge\gamma_{2\;CB}+\sigma^{AB}_{1}\wedge\Gamma_{0\;CA}\wedge\gamma_{2\;B}^{C}\right. \\
\left. -\frac{\Lambda}{3}\sigma_{1}^{AB}\wedge\sigma_{2\;AB}-\sigma^{AB}_{1}\wedge\psi_{2\;ABCD}\Sigma_{0}^{CD}\right).
\end{multline*}
The last term vanishes due to the linearized field equation \eqref{LFE3} and the fact that $\psi_{ABCD}=\psi_{(ABCD)}$, while the second and third terms cancel after restructuring the spinor indices.  

All that remains is to check that \eqref{MHV1} has the correct gauge invariance: if one of the ASD states is pure gauge, the amplitude must vanish.  Suppose that $h_{1}$ is pure gauge: $\psi_{1\;ABCD}=0$.  By \eqref{eqn: SDLFE}, we know that $\frac{\Lambda}{3}\sigma_{1}^{AB}=\D_{0}\gamma_{1}^{AB}$, and integrating by parts in \eqref{MHV1} gives
\begin{equation*}
I^{\mathrm{GR}}[1^{-},2^{-},M^{+}]|_{\psi_{1}=0} =\frac{i}{\kappa^{2}}\int_{M}\left(\Sigma_{0}^{AB}\wedge\gamma_{1\;A}^{C}\wedge\gamma_{2\;CB}+\gamma_{1}^{AB}\wedge\D_{0}\sigma_{2\;AB}\right)-\int_{\partial M}\gamma_{1}^{AB}\wedge\sigma_{2\;AB}.
\end{equation*}
The boundary term vanishes at $\scri^{+}$ since $h_{2}|_{\scri^{+}}\in V^{-}$, and also at $\scri^{-}$ since $h_{1}$ is pure gauge.  This leaves us with the bulk terms, which can be evaluated using the linearized field equation \eqref{LFE1} for $h_{2}$:
\begin{multline*}
\int_{M}\left(\Sigma_{0}^{AB}\wedge\gamma_{1\;A}^{C}\wedge\gamma_{2\;CB}+\gamma_{1}^{AB}\wedge\D_{0}\sigma_{2\;AB}\right) \\
=\int_{M}\left(\Sigma_{0}^{AB}\wedge\gamma_{1\;A}^{C}\wedge\gamma_{2\;CB}-2\gamma_{1}^{AB}\wedge\gamma_{2\;C(A}\wedge\Sigma_{0\;B)}^{C}\right) =0,
\end{multline*}
with the final equality following after re-arranging contractions on spinor indices.     $\Box$

\medskip

The final step is to obtain the conformal/Einstein gravity correspondence for this generating functional.  Upon restricting to Einstein scattering states, it is obvious that the generating functional in conformal gravity with two negative helicity gravitons and a SD background is given by the second term in \eqref{CGA3}:
\be{CGGF*}
I^{\mathrm{CG}}[1^{-},2^{-},M^{+}]=\frac{2i}{\varepsilon^{2}}\int_{M} \d\mu \; \psi_{1}^{ABCD}\psi_{2\;ABCD},
\ee
where $M$ is again the SD background which encodes the $n-2$ remaining gravitons.  By the conformal/Einstein gravity correspondence, we should be able to relate $I^{\mathrm{CG}}$ to $I^{\mathrm{GR}}$ on-shell (i.e., by apply the field equations of general relativity), and this is indeed the case \cite{Adamo:2013}.

\begin{propn}\label{CGDS}
On-shell, $I^{\mathrm{GR}}[1^{-},2^{-},M^{+}]=-\frac{3\varepsilon^{2}}{\Lambda\kappa^{2}}I^{\mathrm{CG}}[1^{-},2^{-},M^{+}]$.
\end{propn}
\proof  \eqref{CGGF*} is equivalent to
\begin{equation*}
I^{\mathrm{CG}}[1^{-},2^{-},M^{+}]=\frac{i}{\varepsilon^2}\int_{M}\psi_{1}^{ABCD}\Sigma_{0\;CD}\wedge\psi_{2\;ABEF}\Sigma_{0}^{EF}.
\end{equation*}
Using the linearized field equation \eqref{LFE2} for $h_{2}$, this becomes
\begin{equation*}
\frac{i}{\varepsilon^2}\int_{M}\psi_{1}^{ABCD}\Sigma_{0\;CD}\wedge\left(\D_{0}\gamma_{2\;AB}-\frac{\Lambda}{3}\sigma_{2\;AB}\right).
\end{equation*}
Integrating by parts in the first term gives
\begin{equation*}
-\int_{M}\D_{0}\psi_{1}^{ABCD}\Sigma_{0\;CD}\wedge\gamma_{2\;AB}+\int_{\partial M}\psi_{1}^{ABCD}\Sigma_{0\;CD}\wedge\gamma_{2\;AB}=\int_{\partial M}\psi_{1}^{ABCD}\Sigma_{0\;CD}\wedge\gamma_{2\;AB},
\end{equation*}
since $\psi_{1}$ is a linearized Weyl spinor.  In the second term, a combination of both field equations \eqref{LFE2} for $h_{1}$ and \eqref{LFE1} for $h_{2}$ as well as integration by parts leaves
\begin{equation*}
-\frac{2\Lambda}{3}\int_{M}\gamma_{1}^{AB}\wedge\gamma_{2\;C(A}\wedge\Sigma_{0\;B)}^{C}+\frac{\Lambda^{2}}{9}\int_{M}\sigma_{1}^{AB}\wedge\sigma_{2\;AB}-\frac{\Lambda}{3}\int_{\partial M}\gamma_{1}^{AB}\wedge\sigma_{2\;AB}.
\end{equation*}
Combining both terms gives:
\begin{multline*}
I^{\mathrm{CG}}[1^{-},2^{-},M^{+}]=\frac{i}{\varepsilon^2}\left(-\frac{2\Lambda}{3}\int_{M}\gamma_{1}^{AB}\wedge\gamma_{2\;C(A}\wedge\Sigma_{0\;B)}^{C}+\frac{\Lambda^{2}}{9}\int_{M}\sigma_{1}^{AB}\wedge\sigma_{2\;AB} \right) \\
-\frac{i}{\varepsilon^2}\left(\int_{\partial M}\psi_{1}^{ABCD}\Sigma_{0\;CD}\wedge\gamma_{2\;AB}-\frac{\Lambda}{3}\int_{\partial M}\gamma^{AB}_{1}\wedge\sigma_{2\;AB}\right) \\
=-\frac{\Lambda \kappa^{2}}{3\varepsilon^{2}}I^{\mathrm{GR}}[1^{-},2^{-},M^{+}] +\mbox{boundary terms}.
\end{multline*}

The proof is complete if we can show that the boundary terms vanish.  Applying \eqref{LFE2} to the first of these terms leaves us
\begin{equation*}
\mbox{boundary terms} \sim \int_{\partial M}\D_{0}\gamma_{1}^{AB}\wedge\gamma_{2\;AB}-\frac{\Lambda}{3}\int_{\partial M}\gamma_{2}^{AB}\wedge\sigma_{1\;AB}-\frac{\Lambda}{3}\int_{\partial M}\gamma_{1}^{AB}\wedge\sigma_{2\;AB},
\end{equation*}
with the second and third terms cancelling due to skew symmetry in $h_{1},h_{2}$.  Finally, 
\begin{multline*}
\int_{\partial M}\D_{0}\gamma_{1}^{AB}\wedge\gamma_{2\;AB}=\int_{\scri^{+}}\D_{0}\gamma_{1}^{AB}\wedge\gamma_{2\;AB}-\int_{\scri^{-}}\D_{0}\gamma_{1}^{AB}\wedge \gamma_{2\;AB} \\
=-\int_{\scri^{+}}\gamma_{1}^{AB}\wedge\D_{0}\gamma_{2\;AB}-\int_{\scri^{-}}\D_{0}\gamma_{1}^{AB}\wedge\gamma_{2\;AB}=0,
\end{multline*}
by the fact that $h_{1}|_{\scri^{-}}\in V^{-}$ and $h_{2}|_{\scri^{+}}\in V^{-}$, as required.     $\Box$

\medskip

Note that the result of this proposition is in precise agreement with the prefactors predicted by Anderson's theorem in \eqref{CGB2}.


\subsection{Minimal and non-minimal conformal super-gravity}\label{minimal}

It is natural to ask if the embedding of  Einstein gravity into conformal gravity persists in the presence of supersymmetry.  Analogues of conformal gravity with extended supersymmetry were first constructed in \cite{Bergshoeff:1980is}, and it is believed that these theories are well-defined for $\cN\leq 4$ (c.f., \cite{Ferrara:1977ij, deWit:1978pd}).  In this paper, we are concerned primarily with $\cN=4$ conformal supergravities (CSGs), since this is the degree of supersymmetry that arises most naturally in twistor theory.  This $\cN=4$ CSG comes in two basic phenotypes: \emph{minimal} and \emph{non-minimal} based upon the presence of a certain global symmetry.  The non-minimal type depends essentially on a free function of one variable.  Einstein supergravity embeds into minimal CSG, but \emph{not} into the non-minimal models.

The field content of $\cN=4$ CSG consists of the spin-2 conformal gravitons along with bosonic fields $V^{a}_{\mu\;b}$, anti-self-dual tensors $T^{ab}_{\mu\nu}$, scalars $\{E_{ab}, D^{ab}_{cd}, \varphi\}$ and fermions $\{\psi^{a}_{\mu}, \chi^{a}_{bc}, \lambda_{a}\}$, where $a=1,\ldots,4$ is a $\SU(4)$ $R$-symmetry index.  \emph{Minimal} $\cN=4$ CSG is characterized by a global $\SU(1,1)$ symmetry acting non-linearly on the complex scalar $\varphi$ (essentially the action of $\SU(1,1)$ on the upper-half plane) \cite{Bergshoeff:1980is}.  This relates to the presence of $\cN=4$ Poincar\'e supergravity sitting inside the CSG \cite{Cremmer:1977tt}.  The minimal model also has a degenerate limit where $\SU(1,1)$ is replaced by a linear $\E_{2}$ action (the Euclidean symmetries of the plane); once again this has an analogue in $\cN=4$ Einstein supergravity, and also arises in coupling $\cN=1$ supergravity to a scalar multiplet \cite{Ferrara:1976ni, Ferrara:1976kg, Cremmer:1977tt}. 
  

A general conformally invariant theory of gravity has a Lagrangian of the form
$$
\cL= f(\varphi) \Psi^2 + \varphi \Box^2\bar\varphi+ c.c. +\ldots\, ,
$$
where we just give two indicative terms of a rather extended Lagrangian.  Because the field $\varphi$ has conformal weight zero, we are allowed an arbitrary function $f(\varphi)$ as a coefficient of the self-dual Weyl tensor squared $\Psi^2$.  This will have a supersymmetric extension for arbitrary analytic $f$.    

\begin{figure}
\centering
\includegraphics[width=3.25 in, height=1.25 in]{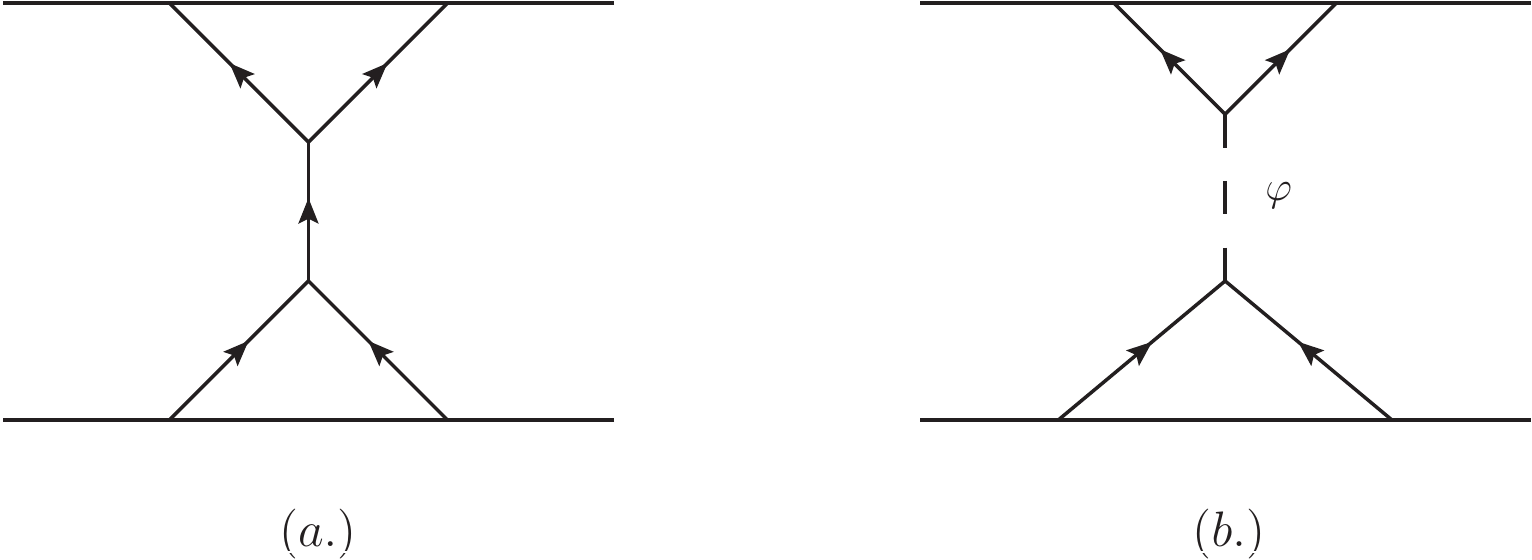}\caption{\textit{In minimal} $\cN=4$ \textit{CSG, couplings between gravitons and the scalar $\varphi$ are excluded in the bulk} (\emph{a}.); \textit{in the non-minimal model they are not} (\emph{b}.).}\label{CSGs}
\end{figure}

In the minimal $\cN=4$ case, the aforementioned $\SU(1,1)$ symmetry leads to a unique $\cN=4$ CSG Lagrangian. It follows from symmetry under the $\U(1)$ subgroup of $\SU(1,1)$ that we must have $f\equiv 1$, giving the Lagrangian:
\begin{equation*}
\cL^{\mathrm{min}}=C^{\mu\nu\rho\sigma}C_{\mu\nu\rho\sigma}+\varphi \Box^{2}\bar{\varphi} +\cdots\, .
\end{equation*}
Einstein supergravities at $\cN=4$ can be constructed from minimal CSG \cite{deRoo:1985jh} and so restricting to Einstein scattering states, Maldacena's argument should still apply and we can extract the tree-level Einstein gravity scattering amplitudes (see Figure \ref{CSGs} (\emph{a})).   

Without the global $\SU(1,1)$ symmetry, there are no constraints on $f(\varphi)$, which leads to couplings between the complex scalar $\varphi$ and the Weyl curvature.  Such $\cN=4$ CSG theories are referred to as \emph{non-minimal}, and were first conjectured to exist in \cite{Fradkin:1983tg, Fradkin:1985am}.  If $f'\neq 0$, the Weyl tensor will provide a source for the scalar field and \emph{vice versa}, so even if $\varphi$ vanishes asymptotically it will become nontrivial in the interior.  Einstein gravity will not be a subset of this theory and there will in general be no embedding of Einstein solutions into non-minimal CSG.

At the level of scattering amplitudes, conformal graviton scattering states in the non-minimal theory can interact with the scalar in the bulk via three-point vertices of the form $\varphi (\mbox{Weyl})^{2}$.  This means that a tree-level scattering amplitude for conformal gravitons will include Feynman diagrams for which there is no analogue in Einstein supergravity, as illustrated in Figure \ref{CSGs} (\emph{b}).  Without a consistent algorithm for subtracting these diagrams, Maldacena's argument \emph{can not} be applied to non-minimal CSG.  The theory arising from the Berkovits-Witten twistor string is understood to be an example of non-minimal CSG, with $f(\varphi)=\e^\varphi$ \cite{Berkovits:2004jj}.  And indeed, spurious amplitudes related to the non-minimal coupling between conformal gravitons and scalars were found explicitly in \cite{Dolan:2008gc, Adamo:2012nn}.

While there is some doubt over whether non-minimal CSG is well-defined at the quantum level \cite{Romer:1985yg, Buchbinder:2012uh}, minimal conformal gravity maintains some independent interest.  It has been shown that minimal $\cN=4$ CSG interacting with a $\SU(2)\times\U(1)$ $\cN=4$ SYM theory is ultraviolet finite and power-counting renormalizable \cite{Fradkin:1981jc, Fradkin:1985am}.  This theory can be obtained as a gauge theory of the superconformal group $\SU(2,2|4)$.  A weaker version of the minimal Lagrangian can also be obtained by coupling abelian $\cN=4$ SYM to a $\cN=4$ CSG background \cite{deRoo:1984gd, deRoo:1985jh} and extracting the UV divergent portion of the partition function \cite{Liu:1998bu, Buchbinder:2012uh}.  The theory has even been proposed as a basic model for quantum gravity (c.f., \cite{Mannheim:2009qi, Mannheim:2011ds}).




\subsection{Further remarks on conformal gravity polarization states}\label{pol-states}

We include the following in order to make contact with standard calculations.  Standard momentum eigenstates for spin-two fields with 4-momentum $k_{AA'}=p_{A}\tilde{p}_{A'}$ are given by
\begin{equation*}
\psi_{ABCD}=p_Ap_Bp_Cp_D \e^{ik\cdot x}\, , 
\end{equation*}
where the polarization information is contained in the choice of scale of $p_A$. 

As conformal gravity has fourth-order equations of motion, we need more polarization states and--as mentioned above--it is usually thought that twice as many suffice \cite{Berkovits:2004jj, Mannheim:2009qi}, although we will present three here to line up with the counting from twistor space.  The first two arise from \eqref{spin-two-E} as the pair 
\be{CG-pol}
\Psi_{ABCD}=p_Ap_Bp_Cp_D \e^{ik\cdot x}
\, , \qquad \Psi'_{ABCD}=x^2 p_Ap_Bp_Cp_D \e^{ik\cdot x}\, ,
\ee
and similarly for $\tilde{\Psi}$, $\tilde{\Psi}'$.  This framework can be used to characterize Einstein polarization states inside conformal gravity by taking the linear combination of conformal gravity polarization states which vanishes at the hypersurface at infinity.  On the affine patch of metric \eqref{dSmetric2} we have
\be{Einstein-pol}
\Psi^\Lambda_{ABCD}=(1-\Lambda x^2)p_Ap_Bp_Cp_D \e^{ik\cdot x}\, ,
\ee
while on the Poincar\'{e} patch of de Sitter space \eqref{dSmetric3}, this is  
\be{Einstein-pol2}
\Psi^\Lambda_{ABCD}= \sqrt{\frac{\Lambda}{3}} \;t\; p_Ap_Bp_Cp_D \e^{ik\cdot x}\, .
\ee

Another linearized conformal gravity solution that is missed by the above is
\be{Missing}
\Psi_{ABCD}=\alpha_{(A}p_Bp_Cp_{D)}\e^{ik\cdot x}\,  ,
\ee
where $\alpha_{A}$ is an arbitrary constant spinor.  The general solution for the spin two equation can be expressed by Fourier transform as 
$$
\psi_{ABCD}(x)=\int \rd^4 k \; \delta(k^2)\;  \psi(k)\; p_Ap_Bp_Cp_D \e^{ik\cdot x}\, .
$$
Similarly, the general solution to the linearized Bach equations can be expressed as
\be{gen-soln}
\Psi_{ABCD}(x)=\int \rd^4 k \;\left(\delta(k^2)  \Psi_{0}(k)_{(A} + p_{(A}\delta'(k^2) \Psi_1(k)\right)\; p_Bp_Cp_{D)} \e^{ik\cdot x}\, .
\ee
This can be seen by taking the Fourier transform of the Bach equations to yield
$$
k^{AA'}k^{BB'}\Psi_{ABCD}(k)=0.
$$
Multiplying by $k$ twice more we discover that $(k^2)^2\Psi_{ABCD}(k)=0$ so that 
$$\Psi_{ABCD}(k)=\Psi_{0\;ABCD}\;\delta(k^2)+\Psi_{1\;ABCD}\;\delta'(k^2).$$
Introduce $p_A=k_{AA'}o^{A'}$ so that $k^{AA'}p_A=\frac{k^2}{2} o^{A'}$.  Then it is straightforward to see that the field equations are satisfied by \eqref{gen-soln} and that this is the general solution.  Integrating by parts in \eqref{gen-soln}, we can eliminate $\delta'(k^2)$ in favour of $\delta(k^2)$ but will then pick up explicit dependence on $x^\mu$ making contact with the polarization states \eqref{CG-pol}.


\subsection{Proof of Lemma \ref{taucont}}
\label{taucontapp}
 
In this appendix, we provide the proof of lemma \ref{taucont} from the text, regarding propagators involving the contact structure $\tau$.

\begin{lemma}
A Feynman diagram containing an \emph{isolated} propagator connecting a white vertex to a grey vertex vanishes following integration by parts.  In particular, an isolated propagator between a white vertex $i$ and a grey vertex $1$ leads to a factor 
\begin{equation*}
2\Lambda\D\sigma_{1}\;\sigma_{1A}\int_{\CP^1}\d_{i}\left(\frac{\sigma_{i}^{A}(\xi 1)h_{i}}{(1i)^2 (\xi i)}\right),
\end{equation*}
where $\d_{i}$ is the exterior derivative in $\sigma_{i}$.
\end{lemma}

\proof  Consider a diagram with a disconnected component with a single
\emph{isolated} arrow from (white) vertex $i$ to $\tau_{1}$.  
Using the fact that the propagator  replaces $Z(x,\sigma_1)$ according to
\be{Picit1A}
Z^{I}(x,\sigma_1)\rightarrow\int_{\CP^{1}}\frac{\D\sigma_i}{(1
  i)}\frac{(\xi 1)^2}{(\xi i)^2}I^{IJ}\partial_J h_i,
\ee
we see that this disconnected component corresponds to a factor
\begin{multline*}
 I_{IJ}\int_{\CP^{1}}\frac{\D\sigma_{i}}{(1i)^2}\frac{(\xi 1)}{(\xi i)^2}\left[(\xi 1)(1i) \rd Z^{J}_1 I^{IK}\p_K h_i+ \left((\xi 1)(i\d\sigma_{1})+ 2 (1i)(\xi\d\sigma_{1})\right) Z^{I}_1I^{JK}\p_Kh_i\right] \\
=  I_{IJ}\int_{\CP^{1}}\frac{\D\sigma_{i}}{(1i)^2}\frac{(\xi 1)}{(\xi
  i)^2}\left[(\xi 1)(1i) \rd Z^{J}_1 I^{IK}\p_Kh_i +\left(\D\sigma_{1}
  (\xi i) 
 +(1i)(\xi\d\sigma_{1})\right)
Z^{I}_1I^{JK}\p_Kh_i\right],
\end{multline*}
with the second expression following by the Schouten identity.
The map to twistor space takes the form $Z^{I}(\sigma)=X^{I}_{A}\sigma^{A}=(X\sigma)^{I}$, so the Schouten identity gives
\begin{equation*}
\rd Z^{J}(\sigma_{1})\;(1i)=Z^{J}(\sigma_{i})\;\D\sigma_{1} - Z^{J}(\sigma_{1})\;(i\d\sigma_{1}),
\end{equation*}
and feeding this into the above expression leaves us with
\begin{equation*}
I_{IJ} \D\sigma_{1}\int_{\CP^1}\frac{\D\sigma_{i}}{(1i)^2}\frac{(\xi
  1)}{(\xi i)^2}\left(2(\xi i) \; Z^{I}_1-(\xi 1)
  \;Z^{I}_i\right)I^{JK}\p_Kh_i\, .
\end{equation*}
Using $I_{IJ}I^{JK}=\Lambda \delta_I^K$ we have 
\begin{multline*}
\Lambda\D\sigma_{1} \int_{\CP^{1}}\frac{\D\sigma_{i}\;(\xi 1)}{(1i)^{2}(\xi i)^{2}}\left(2(\xi i)Z^I_1-(\xi 1)Z^{I}_{i}\right)\partial_{I} h_i \\
=2\Lambda \D\sigma_{1} \int_{\CP^1}\frac{\D\sigma_{i}\;(\xi 1)}{(1i)^{2}(\xi i)^{2}}\left((\xi i)\sigma_{1}\cdot\partial_{i} h_i-(\xi 1)h_{i}\right) \\
=2\Lambda\D\sigma_{1} \sigma_{1A}\int_{\CP^1}\frac{\partial}{\partial\sigma_{iA}}\left(\frac{\D\sigma_{i}\;(\xi 1)h_{i}}{(1i)^{2}(\xi i)}\right) 
= 2\Lambda \D\sigma_{1} \sigma_{1A}\int_{\CP^1}\partial_{i} \left(\frac{\sigma^{A}_{i}(\xi 1)h_{i}}{(1i)^{2}(\xi i)}\right).
\end{multline*}
In the second line we have used the homogeneity relation, chain rule, and the linearity of $Z_i$ in $\sigma_{i}$ to deduce that $\sigma_{1}\cdot \p_{i}h(Z(\sigma_{i}))=Z^I(\sigma_1)\p_I h(Z(\sigma_i))$.  

The integrand of this expression has potential poles at $\sigma_{i}=\sigma_{1}, \xi$ which could lead to boundary contributions when we apply Stokes theorem.  If we take $\sigma_{i}=\sigma_{1}+z\xi$, then the integral takes the form:
\begin{equation*}
\int_{\CP^1}\partial_{i} \left(\frac{g(z)\d\bar{z}}{z}\right)=\oint_{r=\infty}\frac{g(z)\d\bar{z}}{z}-\oint_{r=0}\frac{g(z)\d\bar{z}}{z},
\end{equation*}
where $g(z)$ is a smooth weighted holomorphic function.  Writing $z=re^{i\theta}$, we are left with
\begin{equation*}
-i\oint_{r=\infty}g(z)e^{-2i\theta}\d\theta +i\oint_{r=0}g(z)e^{-2i\theta}\d\theta =0,
\end{equation*}
so any potential boundary terms do indeed vanish. The case with two isolated contractions into $\tau_{1}$ from vertices $i$ and $j$ follows similarly.     $\Box$


\subsection{Independence of the Reference Spinor}
\label{XI}   

In this appendix we explicitly compute the infinitesimal variation $\d_{\xi}\cM_{n,0}$.  This is easiest if we use the representation of $\cM_{n,0}$ given by \eqref{MHVamp}; the proof of $\xi$-independence can also be accomplished using \eqref{MHVamp2}, but requires a bit more finesse. 

We can compute the variation directly from \eqref{MHVamp} by using the basic property of determinants: $\d_{\xi}|\HH|=\tr[\mathrm{adj}(\HH)\d_{\xi}\HH]$.  This leads to (ignoring irrelevant overall factors):
\begin{multline}\label{xivar1}
\d_{\xi}\cM_{n,0}=\int_{\CM_{n,1}}\d^{4|8}x\;\left[\sum_{i}\left|\HH^{12i}_{12i}\right|\left((X^2)^2 \; \d_{\xi}\HH_{ii}+X^{2}\;\d_{\xi}\psi^{1}_{i}\right) \right. \\
\left. +\sum_{i,j}\left|\HH^{12ij}_{12ij}\right|\left(X^{2}\;\psi^{1}_{i}\;\d_{\xi}\HH_{jj}+X^{2}\;\d_{\xi}\omega^{1}_{ij}+\d_{\xi}\psi^{1}_{i}\;\psi^{2}_{j}+\psi^{1}_{i}\;\d_{\xi}\psi^{2}_{j}\right)\right. \\
\left. +\sum_{i,j,k}\left|\HH^{12ijk}_{12ijk}\right| \left(X^{2}\;\omega^{1}_{ij}\;\d_{\xi}\HH_{kk}+\psi^{1}_{i}\;\psi^{2}_{j}\;\d_{\xi}\HH_{kk}+\d_{\xi}\psi^{1}_{i}\;\omega^{2}_{jk}+\psi^{1}_{i}\;\d_{\xi}\omega^{2}_{jk}\right)\right. \\
\left. +\sum_{i,j,k,l}\left|\HH^{12ijkl}_{12ijkl}\right| \left(\psi^{1}_{i}\;\omega^{2}_{jk}\;\d_{\xi}\HH_{ll}+\d_{\xi}\omega^{1}_{ij}\;\omega^{2}_{kl}+\omega^{1}_{ij}\;\d_{\xi}\omega^{2}_{kl}\right) \right. \\
\left. +\sum_{i,j,k,l,m}\left|\HH^{12ijklm}_{12ijklm}\right|\;\omega^{1}_{ij}\;\omega^{2}_{kl}\;\d_{\xi}\HH_{mm}\right]\prod_{s=1}^{n}h_{s}\;\D\sigma_{s} \:+(1\leftrightarrow 2).
\end{multline}

To study $\d_{\xi}\cM_{n,0}$, we need the individual variations which appear in \eqref{xivar1}.  These are easily obtained by working with the dual twistor wavefunctions \eqref{dtwf}; after a bit of algebra (including the Schouten identity) we find
\be{hvar}
\d_{\xi}\HH_{ii}=2\sum_{j=1}^{n}\frac{[W_{i},W_{j}](j\xi)}{(i\xi)^{3}}\D\xi = 2\frac{[W_{i},\cP\cdot\xi ]}{(i\xi)^3}\D\xi,
\ee
\be{psivar}
\d_{\xi}\psi^{1}_{i}=2i\;\Lambda\frac{Z(\xi)\cdot W_{i}}{(i\xi)^3}\D\xi, \qquad \d_{\xi}\omega^{1}_{ij}=2\Lambda\frac{[W_{i},W_{j}](ij)(1\xi)^{3}}{(1i)^{2}(1j)^{2}(i\xi)^{3}(j\xi)^{3}}\left[(j\xi)(i1)+(i\xi)(j1)\right]\D\xi.
\ee
We will now use these facts to show that $\d_{\xi}\cM_{n,0}$ is a total divergence with respect to the coordinates $X^{JA}$, and hence vanishes.

We can proceed order-by-order with respect to the sums appearing in \eqref{xivar1}.  For instance, the integrand of the first line is
\begin{equation*}
-2i\sum_{i}\left|\HH^{12i}_{12i}\right|\left(i(X^2)^{2}\frac{[W_{i},\cP\cdot\xi ]}{(i\xi)^3}-2\Lambda\;X^{2}\frac{Z(\xi)\cdot W_{i}}{(i\xi)^3}\right) e^{i\cP\cdot X}.
\end{equation*}
But upon inspection, this takes the form of a total divergence:
\be{3ovar}
-2i\frac{\partial}{\partial X^{JA}}\left[(X^2)^{2}\;e^{i\cP\cdot X}\sum_{i}\left|\HH^{12i}_{12i}\right|\;I^{IJ}\frac{W_{i\;I}\xi^{A}}{(i\xi)^{3}}\right]
\ee
The key observation is that (for all terms contributing to $\d_{\xi}\cM_{n,0}$) $X$-dependence only appears through explicit powers of $X^{2}$, the wavefunction factor of $e^{i\cP\cdot X}$, $\psi^{1}_{i}$, or $\d_{\xi}\psi^{1}_{i}$.  Applying this philosophy to the rest of \eqref{xivar1}, we can show that line-by-line it is equal to a total divergence.

If we refer to the contribution of \eqref{3ovar} as the `third-order' contribution (counting the number of rows and columns missing from the determinant factor), then divergences at each order are given as follows:  At fourth-order,
\be{4ovar}
-2i\frac{\partial}{\partial X^{JA}}\left[X^{2}\;e^{i\cP\cdot X}\sum_{i,j}\left|\HH^{12ij}_{12ij}\right|\;\psi^{1}_{i}\;I^{IJ}\frac{W_{j\;I}\xi^{A}}{(j\xi)^{3}}\right] \: +(1\leftrightarrow 2).
\ee
At fifth-order:
\be{5ovar}
-2i\frac{\partial}{\partial X^{JA}}\left[e^{i\cP\cdot X}\sum_{i,j,k}\left|\HH^{12ijk}_{12ijk}\right|\left(X^{2}\omega^{1}_{ij}-\psi^{1}_{i}\psi^{2}_{j}\right)I^{IJ}\frac{W_{k\;I}\xi^{A}}{(k\xi)^{3}}\right] \: (1\leftrightarrow 2).
\ee
At sixth-order:
\be{6ovar}
-2i\frac{\partial}{\partial X^{JA}}\left[e^{i\cP\cdot X}\sum_{i,j,k,l}\left|\HH^{12ijkl}_{12ijkl}\right|\;\psi^{1}_{i}\;\omega^{2}_{jk}\;I^{IJ}\frac{W_{l\;I}\xi^{A}}{(l\xi)^3}\right] \:+(1\leftrightarrow 2).
\ee

At seventh-order, we only have a single term:
\begin{equation*}
2\int\d^{4|8}x\;\sum_{i,j,k,l,m}\left|\HH^{12ijklm}_{12ijklm}\right|\;\omega^{1}_{ij}\;\omega^{2}_{kl}\frac{[W_{m},\cP\cdot\xi]}{(m\xi)^3}\;e^{i\cP\cdot X}\d^{2}\sigma.
\end{equation*}
After using the $\GL(2,\C)$-freedom to fix the scale and position of $\sigma_{1}$ and $\sigma_{2}$, we can simply perform the remaining $\d^{8|8}X$ integral (since $\omega^{1,2}_{ij}$ is independent of $X$), leaving:
\be{7ovar}
2\int \d^{2}\sigma\;\delta^{8|8}(\cP)\sum_{i,j,k,l,m}\left|\HH^{12ijklm}_{12ijklm}\right|\;\omega^{1}_{ij}\;\omega^{2}_{kl}\frac{[W_{m},\cP\cdot\xi]}{(m\xi)^3} =0.
\ee
So the seventh-order contribution to $\d_{\xi}\cM_{n,0}$ vanishes simply due to momentum conservation.  Note that in the calculation of each of these divergences, care must be taken to symmetrize over all indices in the summation as well as $(1\leftrightarrow 2)$ in order to get the correct result.

Finally, we can combine \eqref{3ovar}-\eqref{7ovar} to see that
\be{xivar2*}
\d_{\xi}\cM_{n,0}=\int_{\CM_{n,1}}\frac{\d^{8|8}X}{\mathrm{vol}\;\GL(2,\C)}\frac{\partial}{\partial X^{IA}} V^{IA}=0.
\ee
This vanishing occurs because there are no ambiguities with respect to the compactification of the moduli space at degree one, and $V^{IA}$ is smooth with respect to the $X$ coordinates.

\bibliography{deSitter}
\bibliographystyle{JHEP}

\end{document}